**Mycobiome of the Bat White Nose Syndrome Affected Caves and Mines Reveals Diversity of Fungi and Local Adaptation by the Fungal Pathogen *Pseudogymnoascus* (*Geomyces*) *destructans***

Tao Zhang[1,¶], Tanya R. Victor[1,¶], Sunanda S. Rajkumar[1], Xiaojiang Li[1], Joseph C. Okoniewski[2], Alan C. Hicks[2], April D. Davis[3], Kelly Broussard[3], Shannon L. LaDeau[4], Sudha Chaturvedi[1,5*], and Vishnu Chaturvedi[1,5*]

[1]Mycology Laboratory and [3]Rabies Laboratory, Wadsworth Center, New York State Department of Health, Albany, New York, USA; [2]Bureau of Wildlife, New York State Department of Environmental Conservation, Albany, New York, USA; [4]Cary Institute of Ecosystem Studies, Millbrook, New York, USA; [5]Department of Biomedical Sciences, School of Public Health, University at Albany, Albany, New York, USA

**Key Words:** Culture-dependent, Culture-independent, Psychrophile, Diversity Parameters, ITS, LSU, OTU,

**Running Title:** Mycobiome of White Nose Syndrome Sites

[¶]These authors contributed equally to this work. [*]Corresponding authors: Sudha Chaturvedi, sudha.chaturvedi@health.ny.gov; Vishnu Chaturvedi vishnu.chaturvedi@health.ny.gov

**Abstract**
Current investigations of bat White Nose Syndrome (WNS) and the causative fungus *Pseudogymnoascus* (*Geomyces*) *destructans* (*Pd*) are intensely focused on the reasons for the appearance of the disease in the Northeast and its rapid spread in the US and Canada. Urgent steps are still needed for the mitigation or control of *Pd* to save bats. We hypothesized that a focus on fungal community would advance the understanding of ecology and ecosystem processes that are crucial in the disease transmission cycle. This study was conducted in 2010 - 2011 in New York and Vermont using 90 samples from four mines and two caves situated within the epicenter of WNS. We used culture-dependent (CD) and culture-independent (CI) methods to catalogue all fungi ('mycobiome'). CD methods included fungal isolations followed by phenotypic and molecular identifications. CI methods included amplification of DNA extracted from environmental samples with universal fungal primers followed by cloning and sequencing. CD methods yielded 675 fungal isolates and CI method yielded 594 fungal environmental nucleic acid sequences (FENAS). The core mycobiome of WNS comprised of 136 operational taxonomic units (OTUs) recovered in culture and 248 OTUs recovered in clone libraries. The fungal community was diverse across the sites, although a subgroup of dominant cosmopolitan fungi was present. The frequent recovery of *Pd* (18% of samples positive by culture) even in the presence of dominant, cosmopolitan fungal genera suggests some level of local adaptation in WNS-afflicted habitats, while the extensive distribution of *Pd* (48% of samples positive by real-time PCR) suggests an active reservoir of the pathogen at these sites. These findings underscore the need for integrated disease control measures that target both bats and *Pd* in the hibernacula for the control of WNS.

**Introduction**
White Nose Syndrome (WNS) has devastated bat populations in the Northeastern United States for almost a decade [1,2,3]. The disease first appeared in cave hibernacula near Albany, NY, and it has since spread to many other states and several adjoining provinces in Canada. A number of bat species are affected, particularly little brown bats (*Myotis lucifugus*), northern long-eared bats (*M. septentrionalis*), Indiana bats (*M. sodalis*), and tricolored bats (*Perimyotis subflavus*) [4].

*Pseudogymnoascus* (*Geomyces*) *destructans* (Gargas & Blehert) Minnis & Lindner, the etiologic agent of WNS, is a newly recognized fungus [5,6]. It is well adapted to grow under the cold conditions found in caves and mines ('psychrophile'), secretes proteolytic enzymes similar to fungi that cause skin infections ('dermatophyte'), and seems to have a single genetic population in US ('clonal') [5,7,8,9,10]. The closely related *Geomyces pannorum sensu lato* is a widely dispersed species complex of fungi that share phylogenetic, psychrophilic and pathogenic attributes with *Pd* [11,12,13]. *Pseudogymnoascus destructans* infects hibernating bats and is capable of invading living tissue [1,5,14,15,16]. The infection disrupts the normal sleep and arousal cycle of infected bats causing a premature depletion of fat reserves necessary for the completion of hibernation and successful spring emergence. Loss of energy reserves may be compounded by dehydration and electrolyte imbalances related to skin damage as winter proceeds; secondary bacterial infections of fungal lesions may ensue post-emergence [17,18,19]. Historically, *Pd* appears to be quite well established in European caves, but bats harboring the fungus seldom exhibit pathology similar to that seen in the North American bats, and no mass mortality has been recorded in Europe [10,15]. It is unknown whether smaller winter colony densities in Europe or other environmental conditions may explain this discrepancy [9].

There are only few ecological studies focused on *Pd* and other fungi in WNS affected sites. An extensive survey of affected sites revealed wide distribution of *Pd* DNA [20,21,22]. However, DNA based surveys provide a limited assessment of fungal diversity. Contemporary fungal community studies have employed an integrated approach based upon culture-dependent (CD) and culture-independent (CI) methods for the better estimation of true fungal diversity [23,24]. The CI method is crucial considering that more than 90% of the extant fungi are unlikely to be documented by the culture-dependent method alone; CI allows better insight into fungal diversity [25,26,27,28,29]. Therefore, the present study used CD and CI methods to assess *Pd* and fungal diversity ('mycobiome') in some of the worst affected caves and mines in the WNS outbreak [1,2,30]. The broad objective of our work is to advance the understanding of the fungal ecology and ecosystem processes that might be crucial in the transmission of WNS.

## Materials and Methods

### Ethics statement

Landowner permission to collect samples was granted at all six study sites. Sites included one publicly owned property (Hailes Cave in John Boyd Thatcher State Park, NYS Department of Parks, Recreation and Historic Preservation) as well as five sites on private lands. All collecting of samples in New York State (NYS) was done with the permission and cooperation of DEC as supplemental activities during standard winter hibernacula surveys conducted by the NYS Department of Environmental Conservation (NYSDEC). Aeolus Cave samples were collected with the permission and cooperation of the Vermont Department of Fish and Wildlife (VDFW) as a supplemental activity during a scheduled visit to the site. Authors collected samples as members of the agency teams were counting the bats, thus there was no additional disturbance events. No visitation or sampling permits were required in either state. Because these sites are all important bat hibernacula, and states are trying to coordinate all work conducted within their borders relating to bats, the authors worked closely with the bat specialists for the respective State wildlife agencies (Carl Herzog, NYSDEC, 625 Broadway, Albany, NY; Scott Darling, VTFW, 271 North Main Street, Rutland, VT). These individuals will provide permit information or landowner contact information for projects that they deem consistent with the protection of the sites and the bats. Consequently, the site coordinates are not being provided for these study sites to prevent unauthorized access.

**Sites and samples**

The six sites selected for this study are located in what is considered the 'epicenter' of the WNS outbreak, with the last, the Bennett Hill Hitchcock Mine (Hitchcock Mine), being confirmed infected during 2009 [31]. Two sites are solution caves. Aeolus Cave, in Bennington County VT, has no flowing water, while Hailes Cave, in Albany County, NY, has a resurgent stream that seasonally covers the entire width of the cave and on rare occasions, floods most of the cave. Four sample sites were abandoned mines. The Walter Williams Preserve (Williams Preserve) in Ulster County, NY, was a cement mine. The Hitchcock Mine and the Barton Hill Mines in Essex County NY, and Main Graphite Mine (Graphite Mine) in Warren County NY were all iron mines. These six sites host the largest winter colonies of hibernating bats in the Northeast. Pre-WNS counts ranged from 16,134 in Hailes Cave to 185,022 in the Graphite Mine (NYSDEC Files). The population estimate of 300,000 ± 30,000 in Aeolus Cave, where most hibernating bats cannot be accessed, is consistent with observations of the level of bat activity both pre- and post-WNS at other sites where winter survey counts have been conducted [32]. Declines due to WNS have varied considerably among study sites, with the lowest survival rate (1%) noted in Graphite Mine [31]. The most recent winter bat counts by NYSDEC at Barton Hill Mine (51,144) and Williams Preserve (16,961) indicated that they apparently have amongst the highest survival rates of any northeast hibernacula and are today the two largest remaining hibernation colonies known in this region of the country. Barton Mine, William Preserve Mine and Hailes Cave have had stable or increasing numbers of hibernating bats for several years. It is not known if this is due to the survival of residents, or the immigration of bats from other sites.

With the exception of sampling conducted in August 2010 at Aeolus Cave, all sampling was completed during the hibernation period (samples collected November 2010 - February 2011). The samples were collected in two ways. Hard rock surfaces were sampled with sterile polyester-tipped swabs (Puritan Medical Products, LLC, Guilford, ME, USA). Swabs were moistened with sterile water if dry swabbing of dry surfaces yielded little visible residue (each site was swabbed two times). Loaded swabs were placed in sterilized Ziploc bags. Sediment on hibernacula floors and some ledges was collected with sterilized metal laboratory spatulas and placed in 50 mL sterile plastic centrifuge tubes. Samples were transported from the field in coolers containing freezer-packs. The samples were processed within two weeks of laboratory storage at 6°C.

Within hibernacula, sampling effort was focused primarily on sites most likely to yield evidence of *Pd.* Swabbing targets included wall and ceiling surfaces and drill holes within a few centimeters of roosting bats, or similar sites known or likely to have been occupied by hibernating bats pre-WNS. Sediment was collected from floor or ledge areas directly below roosting bats. Such sediments frequently included bat feces and remnants of decomposed bats that were likely to have been WNS casualties. A few samples were collected just outside the entrance of Aeolus Cave, and one sample outside the entrance to Barton Mine. Extensive mortality had occurred at these external locations in previous winters. In total, 90 samples were obtained including 15 from Aeolus Cave, 20 from Barton Mine, 19 from Graphite Mine, 9 from Hailes Cave, 2 samples from Hitchcock Mine, and 25 from Williams

Preserve Mine (Table S1). In few instances, agar plates from four sites of William Preserve Mine were also exposed (5 - 10 min) to recover aerial spores of *Pd*, but this was not the primary focus of the study.

**Fungal diversity by culture dependent methods**
Approximately 100 mg of sediments were mixed with 1.0 mL of sterile distilled water, vortexed vigorously for 2 min in sterile plastic centrifuge tubes, and allowed to stand for 5 min. Two hundred µL of supernatant was spread on 150-mm diameter sterile Petri plates containing Sabouraud dextrose agar fortified with antibacterials (SDA-A) and Rose Bengal Chloramphenicol (RBC) agar, respectively [1][33]. Each swab sample was processed for the fungal recovery by first submerging it into 500 µL of sterile distilled water in sterile plastic centrifuge tube, vortexing vigorously for 2 min and spreading 200 µL aliquots on culture plates as described above for the sediments. The remaining 300 µL of swab suspension was saved for DNA extraction (described in a later section). All plates were incubated at 15°C, checked periodically for fungal growth for up to 45 days, and all colonies transferred to SDA-A plates, and re-incubated at 15°C. The fungal identification was done by morphological, biochemical, and molecular methods [1,34]. For molecular testing, DNA from single colonies was extracted with a rapid thermolysis-phenol extraction method as described in a published protocol [35]. In brief, a loopful of the fungal colony was removed from the agar plate and placed in 300 µL of modified genomic DNA extraction buffer (100 mM Tris [pH 8.0], 10 mM EDTA, 2% SDS, 1.4 M NaCl, 1% CTAB, 0.4 µg/ml proteinase K). The mixture was incubated at 65°C for 1 hr followed by chloroform: isoamyl alcohol (24:1) extraction, and precipitation with isopropanol and washing with 70% ethanol. The precipitated DNA was centrifuged at 12,000 rpm and the resulting pellet was dried under air and dissolved in 50 µL of Tris-EDTA (TE) buffer. The fungal DNA was used for the amplifications of the internal transcribed spacer (ITS) regions 1 and 2 (ITS1, 5.8S, and ITS2) or ITS2 region alone of the ribosomal gene using primer set V1827 5'-GGAAGTAAAAGTCGTAACAAGG-3' or V49 5'-GCATCGATGAAGAACGCAGC-3' and V50 5'-TCCTCCGCTTATTGATATGC-3' and proof reading KlenTaq DNA polymerase (Sigma-Aldrich, St. Louis, MO, USA). In few instances where ITS regions failed to provide fungal identification, the D1/D2 region of the large subunit (LSU) of the 28S rDNA gene was PCR amplified using primer set V1798 5'GCATATCAATAAGCGGAGGAAAAG-3' and V1799 5'-GGTCCGTGTTTCAAGACGG-3'. PCR were carried out as described in published protocols [36,37]. Briefly, the initial denaturation was done at 95°C for 3 min followed by 30 cycles of 94°C for 1 min, annealing at 55°C for 1min and extension at 68°C for 2 min, followed by final extension at 68°C for 10 min. The PCR amplicons were sequenced, assembled, and edited using Sequencher software 4.6 (Gene Codes Corp., Ann Arbor, MI, USA) and BLAST searched against two databases - GenBank (www.ncbi.nlm.nih.gov) and Centraalbureau voor Schimmelcultures (www.cbs.knaw.nl). Operational taxonomic units (OTUs) with a 97% similarity threshold were defined by the average neighbor hierarchical clustering algorithm using Mothur program [38]. Those OTUs that had similarities greater than 97% with GenBank records were defined as known OTUs; those with less than 97% similarities were defined as unknown OTUs. All fungal isolates were stored at -80°C in sterile cryogenic vials containing 15% sterile glycerol; all isolates were catalogued in the culture collection of the Mycology Laboratory, New York State Department of Health, Albany, NY, USA.

**Fungal diversity by culture independent methods**
Approximately 100 mg of sediment samples and remaining swab suspensions (300 µL) were processed for DNA extraction using the SoilMaster™ DNA Extraction Kit (Epicentre, Madison, WI, USA) according to the manufacturer's instructions. DNA samples were stored in aliquots at -20°C. The *Pd* DNA in the samples was determined by a published real-time PCR assay that targeted the L-rhamnosidase gene [20]. In brief, the assay was performed with TaqMan chemistry in an iQ5 real-time detection system (Bio-Rad, Hercules, CA, USA). Two microliters of each extracted DNA was added to 18 µL of real-time PCR mix (Roche Applied Science, Indianapolis, IN, USA). The amplifications were performed in duplicate wells. For each sample analyzed, no DNA template control was always included. The samples that crossed the cycle threshold ($C_T$) within 40 cycles were considered positive for *Pd* DNA. The samples negative for *Pd* DNA were tested for the presence of inhibitors by spiking the same reaction with 1.0 nanogram of *Pd* gDNA [39].

For the construction of clone libraries, 24 environmental DNA samples including sediments and swabs representing all mines and caves were randomly selected based on the presence or absence of *Pd* DNA by real-time PCR assay (Table S2). These DNA samples were subjected to ITS and LSU PCR amplifications using published protocols as described in an earlier section [36,37]. Ten microliter of PCR product was checked on 2% agarose gel. The amplicons were purified from 10 µl PCR product using PrepEase® DNA Clean-Up Kit (USB Corporation, Cleveland, OH, USA). Twenty-four individual PCR amplicons (2 µl) each for ITS or LSU were pooled based on either source (cave or mine) or sample type (sediment or swab) as detailed in Table S2. These amplicons were then cloned into pCR2.1-TOPO plasmid and transformed into One-Shot Competent *Escherichia coli* using TOPO TA cloning kit (Invitrogen, Carlsbad, CA, USA). Thus, there were 12 pools each of LSU and ITS clones (Table S2). Positive bacterial clones were screened according to *α*-complementation on LB agar plates supplemented with 100µg/mL ampicillin, and 5-bromo-4-chloro-3-indolyl-*β*-D-galactopyranoside (Sigma-Aldrich, St. Louis, MO, USA). Approximately 20-30 clones from each library (LSU or ITS) were randomly selected, plasmids DNA extracted using standard protocol, digested with *Eco*RI, and digested DNA was separated on 2% agarose gels to check for inserts. Selected clones harboring inserts were sequenced using universal primers M13 and T7 flanking the cloning site of the vector [1]. A total of 500 LSU clones and 300 ITS clones were selected for nucleotide sequencing at the Wadsworth Center Molecular Genetics Core, Albany, NY, USA. Vector sequence contamination was removed from all fungal sequences using the automated vector trimming function in Sequencher 4.6 (Gene Codes, Ann Arbor, MI, USA). All sequences were analyzed using the rDNA Database (http://www.arb-silva.de/) and sequences containing chimeric artifacts were removed using CHECK_CHIMERA program (http://rdp.cme.msu.edu/). Final sequences were analyzed using GenBank and CBS databases as described earlier for the identification of fungal cultures. Sequences that showed matches to protozoa, algae, insects, and plants were excluded from BLAST search results [40].

**Phylogenetic analysis and taxonomic attributions**
Multiple alignments were obtained using the CLUSTALX 1.81 and MAFFT programs [41] [42]. Operational taxonomic units (OTUs) with a 97% similarity threshold were defined by the average neighbor hierarchical clustering algorithm using the Mothur program [38]. Multiple alignments were created with reference to selected GenBank sequences using BioEdit v7.0.9 [43]. The alignments were used in neighbor-joining (NJ) and maximum parsimony (MP) phylogenetic analyses with 1,000 bootstrap replicates using MEGA 5.1 [44]. Values (in percentage) were shown on all branch nodes supported by >50% of the trees, and the taxonomic status of each OTU was deduced with the assistance of annotations of these downloaded sequences. The MycoBank and UniProt (http://www.uniprot.org/taxonomy) served as the source of taxonomic references for fungal species [45] [46].

**Diversity analyses**
The diversity analyses were performed on data obtained from CI method using bioinformatics software freely available for the academic users [38,42]. The analyses were restricted to CI data as it was expected to include all fungi at a sample site. Distance matrixes were constructed for each sample and the combined data from the alignments by using the Mothur program [38]. OTU richness and diversity estimates were calculated using Mothur program [38]: coverage and rarefaction curves of the number of observed OTUs were constructed for theoretical richness in samples using the nonparametric richness approaches ACE and Chao 1. Shannon (H') and Simpson (D) indices were computed to describe OTU diversity [38]. Difference in community composition among samples was investigated by plotting relative abundances of various taxonomic groups at hierarchically nested taxonomic scales for each sample. Venn and Tree shared modules were carried out to calculate the number of shared OTUs and phylogenetic relationship among the different fungal communities, respectively [38].

**Nucleotide sequence accession numbers**
All nucleotide sequences obtained in this study were deposited in GenBank under accession numbers: JX534602-JX534931, JX545249-JX545310, JX534602-JX534931, JX898533-JX898551, and KC009247-KC009285 (LSU sequences); KC008730-KC009128, KC009286-KC009522, KC993815-KC993831, JX898552-JX898635, and JX675050-JX675217 (ITS sequences).

## Results

**Fungi in culture**
Six hundred seventy-five fungal isolates recovered in pure culture were identified by a combination of classical and molecular methods (Table S3). Of these, molecular identification of 399 isolates were confirmed by full-length ITS sequences, 237 isolates by ITS2 sequences and 39 isolates by LSU sequences. The most frequent isolates belonged to Ascomycota followed by early diverging fungal lineages (EDFL) and Basidiomycota (Fig.1). The most abundant classes in Ascomycota were *Eurotiomycetes*, *Leotiomycetes*, and *Sordariomycetes* followed by *Saccharomycetes*, *Dothideomycetes*, and *Hyphomycetes.* The most dominant genera were *Penicillium* (*Eurotiomycetes*), followed by *Geomyces* (*Leotiomycetes*), *Oidiodendron* (*Leotiomycetes*), and *Kernia* (*Sordariomycetes*). *Mortierella* and *Trichosporon* constituted the most abundant genera in EDFL, and Basidiomycota, respectively. Twenty-two LSU OTUs, 81 ITS OTUs and 57 ITS2 OTUs were identified by CD method (Table S4-S6). However, when all the datasets were compared, 24 OTUs were common and therefore, 136 unique OTUs were identified by CD method. The phylogenetic tree by maximum likelihood method revealed three distinct clades including Ascomycota, Basidiomycota and EDFL (Fig. 2). *Pseudogymnoascus destructans* was recovered in culture from 16 of the 90 (18%) environmental samples from four of the six caves and mines (Table S1). The pathogen was not isolated in culture from Barton Mine and Hitchcock Mine. In contrast, *Pd* DNA was detected by real-time PCR assay from 41 of the 86 (48%) environmental samples from all mines and caves (Table S1).

**Fungi in library clones**
We obtained 703 sequences from 451 LSU clones (average length 462 bp) and 252 ITS clones (average length 400 bp) (Table S3). Of these, 594 cloned sequences were assigned to fungal environmental nucleic acid sequences (FENAS) comprising 353 LSU clones and 241 ITS clones. The majority of fungal sequences recovered belonged to the Ascomycota (40%), followed by EDFL (27%) and Basidiomycota (19%), and relatively smaller numbers from Chytridiomycota and Glomeromycota (Fig. 3). The distributions of Ascomycota and EDFL sequences were different in the LSU and ITS clone libraries (Ascomycota relative frequency = 47% LSU, 26% ITS; EDFL relative frequency = 16% LSU, 47% ITS), respectively. Thus, more Ascomycota fungal sequences were identified in the LSU clone library than the ITS library, which yielded more EDFL. Additionally, the smaller number of Basidiomycota clones recovered in the study showed a much closer distribution between the LSU (17%) and ITS (21%) libraries. One hundred eighty nine LSU OTUs and 73 ITS OTUs were identified and of these 14 OTUs were common to both the datasets and hence, 248 OTUs were identified from cloned libraries (Table S7-S8). Of 248 OTUs, 167 LSU OTUs (89%) and 51 ITS OTUs (70%) were singletons. The phylogenetic tree using ITS dataset revealed that all sequences appeared within expected monophyletic groups (Fig. 4). *Pseudogymnoascus destructans* clones were detected at low frequency (five clones). One clone each originated from Barton Mine (BME_D02-07), and Hailes Cave (HCA_D08-05) while three clones originated from Williams Preserve Mine (WMU_D11-40, WMU_D12-21, and WMU_D12-23).

**Comparison of cultures and clones**
Ascomycota predominated among sequences discovered by both CD and CI methods (Fig. 1 & Fig. 3). Nearly equal proportions of EDFL were recovered by two methods, but Basidiomycota sequences were more numerous in CI method (Fig. 3). LSU clone library had more representatives of Ascomycota and non-fungal sequences while ITS clone library showed higher representation of EDFL (Fig. S1). The CD method detected a higher percentage of singletons (Fig. S1). Among the CD isolates and CI clones, 13 OTUs were common (Table S9); the remaining OTUs were detected by only one of the two methods, and the similarity coefficient (Cs) of the two methods was only 0.127. The relative abundance of the most dominant OTUs recovered from the CD, and CI methods were as high as 13.5% (91 isolates within 675 isolates, CD method) and 10.1% (71 clones within 703 clones, CI method), respectively. *Penicillium polonicum* was most abundant in results by CD methods and *Trichosporon dulcitum* in the results by CI methods.

**Diversity analytics**
Rarefaction curves with positive slopes were obtained for OTUs recognized in LSU and ITS clone libraries constructed for CI analyses (Fig. 5). These findings indicated that our sampling did not reach saturation and thus, CI results reflect under sampling of potential OTUs in caves and mines. Similarly, two different richness estimators (ACE and Chao-1) predicted considerably higher number of OTUs at the sample sites than the observed 189 LSU OTUs and 73 ITS OTUs, respectively (Table S10). Shannon and Simpson alpha diversity indices showed a diverse fungal community, which was most pronounced in LSU clones. A similar pattern emerged when Venn diagrams of OTU distributions at the six sample sites were plotted. Most sites had unique OTUs and as few as 1-3 OTUs overlapped among the sample sites (Fig. 6). Notably, *Mortierella* spp., *Trichosporon* spp., and *Geomyces* spp. were frequently recovered from caves and mines samples.

## Discussion

**Mycobiome of WNS sites**
This report constitutes the first published mycobiome of WNS-affected caves and mines. The integrated survey using culture-dependent and -independent methods allowed the discovery of a great diversity of fungi including possible novel taxa. The dominant members of this mycobiome were ascomycetes commonly found in soil, including cold, nutrient poor conditions. Many of these fungi have enhanced keratinolytic or toxigenic activities- *Doratomyces stemonitis*, *Fusarium merismoides*, *Geomyces pannorum sensu lato*, *Kernia* spp., *Oidiodendron truncatum* [39,47,48]. Other generalist fungi isolated in our study were *Penicillium polonicum* and *Polypaecilum botryoides*, which are found in a wide variety of substrates, including dried meat or fish stored under cold conditions. The human and animal pathogenic species in the mycobiome included agents of skin infections such as *Arthroderma vanbreuseghemii*, *Geomyces pannorum sensu lato* and *Oidiodendron truncatum* [13,39,47]. The prominent members of EDFL and Basidiomycota were *Helicostylum pulchrum* and *Trichosporon* spp., respectively; both are widely distributed including in cold environments [49]. Thus, the dominant fungi in caves and mines in Upstate New York and Vermont are similar to typical soil fungal communities found in the coldest parts of the earth [49,50,51].

A recent review of published studies on fungi recovered in culture from cave environments listed over 500 genera and 1,000 species with most studies reporting between 18-25 species [52]. In general agreement with the current study, most of the fungal species were cosmopolitan ascomycetes prevalent in the temperate parts of the world and belonged to the following genera -*Aspergillus*, *Penicillium*, *Mucor*,

*Fusarium*, *Trichoderma*, *Cladosporium*, *Alternaria*, and *Paecilomyces*. No 'cave-specialist' genera have yet been identified. This pattern of over-representation of common ascomycetes is consistent with the culture-dependent results obtained in the present study.

A few other recent studies are worth recapitulating here to underline the emerging information from mycobiomes of diverse habitats. Orgiazzi and colleagues [53] have defined 'core soil mycobiome' of generalist fungi by identifying over 1,600 ITS OTUs in soil samples from distinct ecosystems. Most of the identified OTUs belonged to 'ubiquitous taxa of generalist fungi.' Using a similar approach, Porras-Alfaro *et al*., [54] described 67-78 soil OTUs from semiarid grasslands. Giordano *et al*. [55] utilized both culture-dependent and -independent methods to describe fungal communities on bark beetles; they found 55 species in culture, 33 OTUs in clone libraries and 3 OTUs common in both analyses. Their results are likely to reflect both beetle-specific OTUs, as well as soil fungi. Thus, there is a great variation in the number of fungal OTUs reported from abiotic habitats. It is reasonable to assume that variation in the number of OTUs reported in these studies is not solely due to different analytical methods. Clearly, additional studies that utilize integrated approaches for the mycobiomes are needed for meaningful comparisons across diverse habitats.

**Mycobiome challenges**
The high proportions of species that were found singly and the limited sampling of six sites imply that the actual diversity of fungi in the WNS-infested environment in upstate NY and VT is much higher than detected in this study. This limitation presents both an opportunity and an obstacle in assembly of the mycobiome of WNS affected sites. Future studies must include enhanced sampling and additional media for the culture-recovery of fungi, and more primer sets for the construction of diverse clone libraries.

It is well known that the culture-dependent methods are biased towards rapidly growing cosmopolitan fungi. Besides the physical competition for space, the growth of fastidious fungi is limited by the temperature and length of the incubation period, the composition of the medium and the aerobic conditions for recovery [56,57,58,59]. Thus, the fungi recovered in culture in this study do not represent all culturable fungi at the WNS sites. Along similar lines, the success of the culture-independent method is conditional to the quality of DNA available for amplification and the universality of the primers used for the construction of clone libraries [55]. All of the available 'universal' primers introduce a degree of bias in the amplification of fungal DNA, which could lead to over- or under-representations of certain OTUs. We have utilized three primer sets targeting ITS and LSU regions of the ribosomal genes that were used in other mycobiome studies. However, these primers sets are known to be not equally efficient in the amplification of ascomycetes, basidiomycetes and EDFL [55,60,61,62,63]. Additional difficulties in the use of cloned libraries come from their dependence on the alignments of database nucleotide sequences to determine OTUs. The assignment of ITS sequences is less problematic as there is a comprehensive nucleotide database for the species level identification of many fungi. Higher taxonomic assignments especially of unknown fungi are more problematic as they rely on LSU sequences that are not as numerous in the databases. It has also been recognized that increasing the availability of reference LSU sequences in the databases could lead to definitive assignments of isolates of common soil fungus *Mortierella* to known species within the genus instead of to new OTUs [64]. Mycobiome studies that incorporate CD and CI methods are certainly valuable, but still face many methodological and database challenges.

### *Pseudogymnoascus destructans*

The results of DNA sampling and culture-dependent and -independent surveys all confirmed that *Pd* was an important constituent of fungal communities of surveyed caves and mines. *Pseudogymnoascus destructans* DNA was detected from approximately half of the environmental samples examined by real-time PCR assay. The pathogen was also recovered in culture from four of the six caves and mines. It was not isolated from Barton Mine and Hitchcock Mine. In the latter instance, there were only two samples tested. However, the results from Barton Mine were intriguing, as twenty environmental samples were tested from this site. Inability to obtain *Pd* in culture could be due to overwhelming presence of other fungi at the two sites or on culture plates used for recovery in the laboratory. *Pseudogymnoascus destructans* clones were detected at low frequency by CI method. This finding could be due to limited clones assessed for each library prepared from sediment or swab samples; alternately, fungal DNA from other taxa predominated amplification process.

Notably, sediments from the floor and the swabs from the walls and the roof were positive for *Pd* by one of the three survey methods used. Thus, there was a widespread distribution of the pathogen at the affected sites. Our findings further suggested the existence of a reservoir of *Pd* in WNS-affected caves and mines. The colonies of *Pd* were recovered from culture plates in the presence of many other species, affirming its abundance in the presence of common fungi, which raised the possibility that this newly arrived pathogen was locally adapted [65,66,67]. Experimental studies will be needed to evaluate the extent of local adaptation of *Pd* in caves and mines. The conditions prevailing in caves and mines would indicate a specialized ecological niche for *Pd*. This finding is notable considering other human and animal pathogenic fungi also have a higher incidence in specialized ecological niches. Thus, *Histoplasma capsulatum* is abundant in caves and poses high risk for spelunkers, very high numbers of *Cryptococcus neoformans* have been reported from dried pigeon droppings, and *Penicillium marneffei* is frequently recovered from feces and viscera of bamboo rats [68,69,70,71,72,73].

Three recently published fungal surveys that used culture-dependent methods have partial bearings on our findings. Lorch *et al*. [21] described a soil survey of WNS sites by documenting 324 fungal isolates identified to genus level including *Geomyces* spp. and *Pd*. The study was an early snapshot of the fungal communities at WNS-affected sites and raised the possibility that *Pd* is widely dispersed among cave fungi. Recent reports by Johnson *et al*. [50] and Vanderwolf *et al*. [74] using direct sampling of bats from WNS-free hibernacula, allowed enumeration of fungi on bats and their surroundings before the arrival of WNS. Johnson *et al*. [50]

reported 53 OTUs from five US hibernacula without specifying species. Vanderwolf *et al*. [74] reported 117 species of psychrophilic, coprophilous and keratinolytic fungi from eight Canadian caves. As expected, neither of these studies found any *Pd* in the US and the Canadian sites, which were considered free of WNS at the time of sampling. Taken together, fungal surveys including the present study have now provided evidence to support the widely held perception among bat biologists that the isolation of *Pd* at a given site correlates with the presence of WNS, while the absence of this fungal pathogen predicts a WNS free location. The present study demonstrates that *Pd* is a member of diverse fungal communities inhabiting WNS-affected caves and mines in NY and VT. The ecological interactions that may limit or regulate *Pd* within these communities are not known. Earlier investigations have documented a clonal population of *Pd* as WNS spread across the United States [7,8]; it is not yet clear if *Pd* is spreading by outcompeting or displacing other fungi and/or microbes in the affected sites.

**Acknowledgments**
This study was supported in part with funds from by the U.S. Fish and Wildlife Service and the National Science Foundation (SC, VC). Three undergraduate volunteers Erica Ferrari, Crystal Gunsh and Danielle Willsey helped in initial processing of samples. DNA sequencing was performed at the Wadsworth Center Molecular Genetics Core. We thank three anonymous reviewers for their critical reading and constructive criticism.

**Authors Contributions**
Conception and design of the work- SC, VC
Acquisition of data, or analysis and interpretation of data – TZ, TV, SSR, XL, JCO, AD, KB, ACH, SLD, SC, VC
Drafting the article or revising it critically for important intellectual content-TZ, JCO, ACH, SLD, SC, VC
Final approval of the version to be published- TZ, TV, SSR, XL, JCO, AD, KB, ACH, SLD, SC, VC

**Figure Legends**

Fig.1. Fungal isolates recovered by culture-dependent methods. (A) Relative distribution of all cultures according to different fungal phyla; (B) Relative distribution of isolates identified by homologies to ITS sequences; (C) Relative distribution of isolates identified by homologies to ITS2 sequences. The zero percent assigned in pie chart represents isolates with less than 1% relative distribution.

Fig. 2. Phylogenetic relationships among ITS phylotypes for OTUs isolated by culture-dependent method. Phylotypes recovered in this study are shown in bold type. Sequence code prefix denotes location. ACV, Aeolus Cave, Bennington County, VT; BME, Barton Mine, Essex County, NY; GMW, Graphite Mine, Warren County, NY; HME, Hitchcock Mine, Essex County, NY; WMU, Williams Preserve Mine, Ulster County, NY; HCA, Hailes Cave, Albany. Blue font denotes the isolated phylotypes, red font denotes the *Geomyces* spp. and *Pseudogymnoascus destructans* and black font denotes the ITS sequences of reference strains retrieved from GenBank. Three topologies are supported by the program Mega 5.1 [44]. The numbers at node indicate the bootstrap percentages of 1,000 resamples.

Fig.3. Phyla distribution in culture-independent (CI) clones: A) Relative proportions of different phyla of all clones; B) Relative proportions of different phyla of LSU clones; C) Relative proportions of different phyla of ITS clones. The zero percent assigned in pie chart represents clones with less than 1% relative distribution.

Fig. 4. Phylogenetic relationships among ITS phylotypes for OTUs recovered by culture-independent investigation. Phylotypes recovered during this study are shown in bold type. Sequence code prefix denotes location. ACV, Aeolus Cave, Bennington County, VT; BME, Barton Mine, Essex County, NY; GMW, Graphite Mine, Warren County, NY; HME, Hitchcock Mine, Essex County, NY; WMU, Williams Preserve Mine, Ulster County, NY; HCA, Hailes Cave, Albany. Blue font denotes the recovered phylotypes; red font denotes the *Geomyces* spp. and *Pseudogymnoascus* spp., and black font denotes the ITS sequences of reference strains retrieved from GenBank database. Three topologies are supported by the program Mega 5.1 [44]. The numbers at node indicate the bootstrap percentages of 1,000 resamples.

Fig. 5. Rarefaction curves of estimated OTUs richness across bat WNS-afflicted niches environmental samples. A) LSU clones rarefaction curves analyses, cutoffs=0.03, 0.05, 0.10, respectively; B) ITS clones rarefaction curves analyses, cutoff=0.03.

Fig. 6. Overlap in fungal community composition across bat WNS-afflicted environmental samples. Venn diagram demonstrates the degree of overlap of OTU similarity among niches. OTUs were defined using 97% genetic similarity cut-off. A) Venn diagram displays the overlapping OTUs among ACV, BME, WMU, and HCA; B) Venn diagram displays the overlapping OTUs among GMW, BME, WMU, and HME. ACV, Aeolus Cave, Bennington County, VT; BME, Barton Mine, Essex County, NY; GMW, Graphite Mine, Warren County, NY; HME, Hitchcock Mine, Essex County, NY; WMU, Williams Preserve Mine, Ulster County, NY; HCA, Hailes Cave, Albany.

**References**
1. Chaturvedi V, Springer DJ, Behr MJ, Ramani R, Li X, et al. (2010) Morphological and molecular characterizations of psychrophilic fungus *Geomyces destructans* from New York bats with White Nose Syndrome (WNS). PLoS One 5: e10783.
2. Blehert DS, Hicks AC, Behr M, Meteyer CU, Berlowski-Zier BM, et al. (2009) Bat white-nose syndrome: an emerging fungal pathogen? Science 323: 227.
3. Blehert DS (2012) Fungal disease and the developing story of bat white-nose syndrome. PLoS Pathog 8: e1002779.
4. Foley J, Clifford D, Castle K, Cryan P, Ostfeld RS (2011) Investigating and managing the rapid emergence of white-nose syndrome, a novel, fatal, infectious disease of hibernating bats. Conserv Biol 25: 223-231.
5. Gargas A, Trest MT, Christensen M, Volk TJ, Blehert DS (2009) *Geomyces destructans* sp. nov. associated with Bat white-nose syndrome. Mycotaxon 108: 147-154.
6. Minnis AM, Lindner DL (2013) Phylogenetic evaluation of *Geomyces* and allies reveals no close relatives of *Pseudogymnoascus destructans*, comb. nov., in bat hibernacula of eastern North America. Fungal Biol 117: 638-649.
7. Ren P, Haman KH, Last LA, Rajkumar SS, Keel MK, et al. (2012) Clonal spread of *Geomyces destructans* among bats, midwestern and southern United States. Emerg Infect Dis 18: 883-885.
8. Rajkumar SS, Li X, Rudd RJ, Okoniewski JC, Xu J, et al. (2011) Clonal genotype of *Geomyces destructans* among bats with White Nose Syndrome, New York, USA. Emerg Infect Dis 17: 1273-1276.
9. Wibbelt G, Kurth A, Hellmann D, Weishaar M, Barlow A, et al. (2010) White-nose syndrome fungus (*Geomyces destructans*) in bats, Europe. Emerg Infect Dis 16: 1237-1243.
10. Puechmaille SJ, Verdeyroux P, Fuller H, Gouilh MA, Bekaert M, et al. (2010) White-nose syndrome fungus (*Geomyces destructans*) in bat, France. Emerg Infect Dis 16: 290-293.
11. Gilichinsky D, Rivkina E, Bakermans C, Shcherbakova V, Petrovskaya L, et al. (2005) Biodiversity of cryopegs in permafrost. FEMS Microbiol Ecol 53: 117-128.
12. Kochkina GA, Ivanushkina NE, Akimov VN, Gilichinskii DA, Ozerskaia SM (2007) Halo- and psychrotolerant *Geomyces* fungi from arctic cryopegs and marine deposits. Mikrobiologiia 76: 39-47.
13. Godinho VM, Furbino LE, Santiago IF, Pellizzari FM, Yokoya NS, et al. (2013) Diversity and bioprospecting of fungal communities associated with endemic and cold-adapted macroalgae in Antarctica. ISME J 7: 1434-1451.
14. Meteyer CU, Buckles EL, Blehert DS, Hicks AC, Green DE, et al. (2009) Histopathologic criteria to confirm white-nose syndrome in bats. J Vet Diagn Invest 21: 411-414.
15. Martinkova N, Backor P, Bartonicka T, Blazkova P, Cerveny J, et al. (2010) Increasing incidence of *Geomyces destructans* fungus in bats from the Czech Republic and Slovakia. PLoS One 5: e13853.
16. Chaturvedi V, Chaturvedi S (2011) Editorial: What is in a name? A proposal to use geomycosis instead of White Nose Syndrome (WNS) to describe bat infection caused by *Geomyces destructans*. Mycopathologia 171: 231-233.

17. Cryan PM, Meteyer CU, Boyles JG, Blehert DS (2010) Wing pathology of white-nose syndrome in bats suggests life-threatening disruption of physiology. BMC Biol 8: 135.
18. Reeder DM, Frank CL, Turner GG, Meteyer CU, Kurta A, et al. (2012) Frequent arousal from hibernation linked to severity of infection and mortality in bats with white-nose syndrome. PLoS One 7: e38920.
19. Warnecke L, Turner JM, Bollinger TK, Misra V, Cryan PM, et al. (2013) Pathophysiology of white-nose syndrome in bats: a mechanistic model linking wing damage to mortality. Biol Lett 9: 20130177.
20. Chaturvedi S, Rudd RJ, Davis A, Victor TR, Li X, et al. (2011) Rapid real-time PCR assay for culture and tissue identification of *Geomyces destructans*: the etiologic agent of bat geomycosis (white nose syndrome). Mycopathologia 172: 247-256.
21. Lorch JM, Muller LK, Russell RE, O'Connor M, Lindner DL, et al. (2013) Distribution and Environmental Persistence of the Causative Agent of White-Nose Syndrome, *Geomyces destructans*, in Bat Hibernacula of the Eastern United States. Appl Environ Microbiol 79: 1293-1301.
22. Muller LK, Lorch JM, Lindner DL, O'Connor M, Gargas A, et al. (2012) Bat white-nose syndrome: a real-time TaqMan polymerase chain reaction test targeting the intergenic spacer region of *Geomyces destructans*. Mycologia 105: 253-259.
23. Zhang Y, Zhang S, Wang M, Bai F, Liu X (2010) High diversity of the fungal community structure in naturally-occurring *Ophiocordyceps sinensis*. PLoS One 5: e15570.
24. Scanlan PD, Marchesi JR (2008) Micro-eukaryotic diversity of the human distal gut microbiota: qualitative assessment using culture-dependent and -independent analysis of faeces. ISME J 2: 1183-1193.
25. Hawksworth DL (2012) Global species numbers of fungi: are tropical studies and molecular approaches contributing to a more robust estimate? Biodiversity and Conservation 21: 2425-2433.
26. Bass D, Thomas AR (2011) Three reasons to re-evaluate fungal diversity 'on Earth and in the ocean'. Fungal Biology Reviews 25: 159-164.
27. Rosling A, Cox F, Cruz-Martinez K, Ihrmark K, Grelet GA, et al. (2011) *Archaeorhizomycetes*: unearthing an ancient class of ubiquitous soil fungi. Science 333: 876-879.
28. Torsvik V, Ovreas L (2002) Microbial diversity and function in soil: from genes to ecosystems. Curr Opin Microbiol 5: 240-245.
29. Dahlberg A (2001) Community ecology of ectomycorrhizal fungi: an advancing interdisciplinary field. New Phytologist 150: 555–562.
30. Frick WF, Pollock JF, Hicks AC, Langwig KE, Reynolds DS, et al. (2010) An emerging disease causes regional population collapse of a common North American bat species. Science 329: 679-682.
31. Turner GG, Reeder DM, and Coleman JTH (2011) A Five-year Assessment of Mortality and Geographic Spread of White-Nose Syndrome in North American Bats, with a Look at the Future. Update of White-Nose Syndrome in Bats. Bat Research News 52: 13-27.
32. Davis WH, Hitchcock HB (1965) Biology and Migration of the Bat, *Myotis lucifugus*, in New England. J Mammal 46: 296-313.

33. Corry JEL, Curtis GDW, Baird RM (1999) Handbook of Culture Media for Food Microbiology. Second Edition ed: Elsevier Progress in Industrial Microbiology. pp. 586.
34. de Hoog GS GJ, Gene J, Figueras MJ., editor (2000) Atlas of Clinical Fungi. 2nd ed. ed: Utrecht, The Netherlands: Centraalbureau voor Schimmelcultures. .
35. Moller EM, Bahnweg G, Sandermann H, Geiger HH (1992) A simple and efficient protocol for isolation of high molecular weight DNA from filamentous fungi, fruit bodies, and infected plant tissues. Nucleic Acids Res 20: 6115-6116.
36. White TJ, T. D. Bruns, S. Lee, and J. W. Taylor (1990) Amplification and direct sequencing of fungal ribosomal RNA genes for phylogenetics. In: M. A. Innis DHG, J. J. Sninsky, and T. J. White (ed.), editor. PCR protocols: a guide to methods and applications. New York, N.Y.: Academic Press Inc., pp. 315–324.
37. O'Donnell K (1993) In: Taylor. DRRJW, editor. The Fungal holomorph: mitotic, meiotic and pleomorphic speciation in fungal systematics: proceedings of an international symposium. Wallingford (UK): CAB International. pp. 225-233.
38. Schloss PD, Westcott SL, Ryabin T, Hall JR, Hartmann M, et al. (2009) Introducing mothur: open-source, platform-independent, community-supported software for describing and comparing microbial communities. Appl Environ Microbiol 75: 7537-7541.
39. Marcus CC, Jonathan C, Sushma N, Sudha C, Vishnu C (2013) Draft Genome Sequence of Human Pathogenic Fungus *Geomyces pannorum Sensu Lato* and Bat White Nose Syndrome Pathogenic *Geomyces* (*Pseudogymnoascus*) *destructans*. Genome Announcements 1(6): e010451.
40. Altschul SF, Madden TL, Schaffer AA, Zhang J, Zhang Z, et al. (1997) Gapped BLAST and PSI-BLAST: a new generation of protein database search programs. Nucleic Acids Res 25: 3389-3402.
41. Thompson JD, Gibson TJ, Plewniak F, Jeanmougin F, Higgins DG (1997) The CLUSTAL_X windows interface: flexible strategies for multiple sequence alignment aided by quality analysis tools. Nucleic Acids Res 25: 4876-4882.
42. Katoh K, Asimenos G, Toh H (2009) Multiple alignment of DNA sequences with MAFFT. Methods Mol Biol 537: 39-64.
43. Hall TA (1999) BioEdit: a user-friendly biological sequence alignment editor and analysis program for Windows 95/98/NT. Nucleic Acids Symposium Series 41: 95-98.
44. Tamura K PD, Peterson N, Stecher G, Nei M, and Kumar S. (2011) MEGA5: Molecular Evolutionary Genetics Analysis using Maximum Likelihood, Evolutionary Distance, and Maximum Parsimony Methods. Mol Biol Evol 28: 2731-2739.
45. Crous P.W.  GW, Stalpers J.A., Robert V. and StegehuisG. (2004) MycoBank: an online initiative to launch mycology into the 21st century. Studies in Mycology 50: 19-22.
46. Kirk PM CP, Minter DW, Stalpers JA (2008) Ainsworth & Bisby's Dictionary of the Fungi. Wallingford: CAB International. UK. 771 p.
47. Michel C, Prakash M, Katsuhiko A, Guy H, Toru O, et al. (2004) Handbook of Industrial Mycology: CRC Press. -1 p.
48. Awad MF, Kraume M (2011) Fungal diversity in activated sludge from membrane bioreactors in Berlin. Can J Microbiol 57: 693-698.

49. Bronwyn M. Kirby, Desiré Barnard, I. Marla Tuffin, Cowan DA (2011) Ecological Distribution of Microorganisms in Terrestrial, Psychrophilic Habitats. Extremophiles Handbook: Springer Japan. pp. 839-863.
50. Johnson LJ, Miller AN, McCleery RA, McClanahan R, Kath JA, et al. (2013) Psychrophilic and psychrotolerant fungi on bats and the presence of *Geomyces* spp. on bat wings prior to the arrival of white nose syndrome. Appl Environ Microbiol 79: 5465-5471.
51. Bruce J, Morris EO (1973) Psychrophilic yeasts isolated from marine fish. Antonie Van Leeuwenhoek 39: 331-339.
52. Vanderwolf K. J. DFM, David Malloch and Graham J. Forbes. (2013) A world review of fungi, yeasts, and slime molds in caves. Int J Speleol 42: 77–96.
53. Orgiazzi A, Lumini E, Nilsson RH, Girlanda M, Vizzini A, et al. (2012) Unravelling soil fungal communities from different Mediterranean land-use backgrounds. PLoS One 7: e34847.
54. Porras-Alfaro A, Herrera J, Natvig DO, Lipinski K, Sinsabaugh RL (2010) Diversity and distribution of soil fungal communities in a semiarid grassland. Mycologia 103: 10-21.
55. Luana Giordano MG, Giovanni Nicolotti, Paolo Gonthie (2012) Characterization of fungal communities associated with the bark beetle *Ips typographus* varies depending on detection method, location, and beetle population levels. Mycological Progress 12: 127–140.
56. Unterseher M, Schnittler M (2009) Dilution-to-extinction cultivation of leaf-inhabiting endophytic fungi in beech (*Fagus sylvatica* L.)-different cultivation techniques influence fungal biodiversity assessment. Mycol Res 113: 645-654.
57. Tristan C, Cécile R, Xavier C, Marie-Laure Desprez-Loustaua, Corinne V (2012) Spatial variability of phyllosphere fungal assemblages: genetic distance predominates over geographic distance in a European beech stand (*Fagus sylvatica*). Fungal Ecol 5: 509-520.
58. Bills GF (1995) Analyses of microfungal diversity from a user's perspective. . Canadian Journal of Botany 73: 33-41.
59. Collado J, Platas G, Paulus B, Bills GF (2007) High-throughput culturing of fungi from plant litter by a dilution-to-extinction technique. FEMS Microbiol Ecol 60: 521-533.
60. Bellemain E, Carlsen T, Brochmann C, Coissac E, Taberlet P, Kauserud H. (2010) ITS as an environmental DNA barcode for fungi: an in silico approach reveals potential PCR biases. BMC Microbiol 10.
61. Drell T, Lillsaar T, Tummeleht L, Simm J, Aaspollu A, et al. (2013) Characterization of the vaginal micro- and mycobiome in asymptomatic reproductive-age Estonian women. PLoS One 8: e54379.
62. Gao Z, Li B, Zheng C, Wang G (2008) Molecular detection of fungal communities in the Hawaiian marine sponges *Suberites zeteki* and *Mycale armata*. Appl Environ Microbiol 74: 6091-6101.
63. Dickie IA (2010) Insidious effects of sequencing errors on perceived diversity in molecular surveys. New Phytol 188: 916-918.
64. Nagy GL, Petkovits T, Kovács GM, Vágvölgyi Cs, Papp T. (2011) Where is the hidden fungal diversity hiding? In dusty herbaria or in the dust out there? New Phytol 191: 789-794.

65. Wang S, O'Brien TR, Pava-Ripoll M, St Leger RJ (2011) Local adaptation of an introduced transgenic insect fungal pathogen due to new beneficial mutations. Proc Natl Acad Sci U S A 108: 20449-20454.
66. Giraud T, Gladieux P, Gavrilets S (2010) Linking the emergence of fungal plant diseases with ecological speciation. Trends Ecol Evol 25: 387-395.
67. Hereford J (2009) A quantitative survey of local adaptation and fitness trade-offs. Am Nat 173: 579-588.
68. Emmons CW (1955) Saprophytic sources of *Cryptococcus neoformans* associated with the pigeon (*Columba livia*). Am J Hyg 62: 227-232.
69. Littman ML, Borok R (1968) Relation of the pigeon to cryptococcosis: natural carrier state, heat resistance and survival of *Cryptococcus neoformans*. Mycopathol Mycol Appl 35: 329-345.
70. Emmons CW (1958) Association of bats with histoplasmosis. Public Health Rep 73: 590-595.
71. Lyon GM, Bravo AV, Espino A, Lindsley MD, Gutierrez RE, et al. (2004) Histoplasmosis associated with exploring a bat-inhabited cave in Costa Rica, 1998-1999. Am J Trop Med Hyg 70: 438-442.
72. Deng ZL, Yun M, Ajello L (1986) Human *Penicilliosis marneffei* and its relation to the bamboo rat (*Rhizomys pruinosus*). J Med Vet Mycol 24: 383-389.
73. Ajello L, Padhye AA, Sukroongreung S, Nilakul CH, Tantimavanic S (1995) Occurrence of *Penicillium marneffei* infections among wild bamboo rats in Thailand. Mycopathologia 131: 1-8.
74. Vanderwolf K. J. DFM, David Malloch and Graham J. Forbes (2013) Ectomycota Associated with Hibernating Bats in Eastern Canadian Caves prior to the Emergence of White-Nose Syndrome. Northeastern Naturalist 20: 115-130.

**Figure 1.**

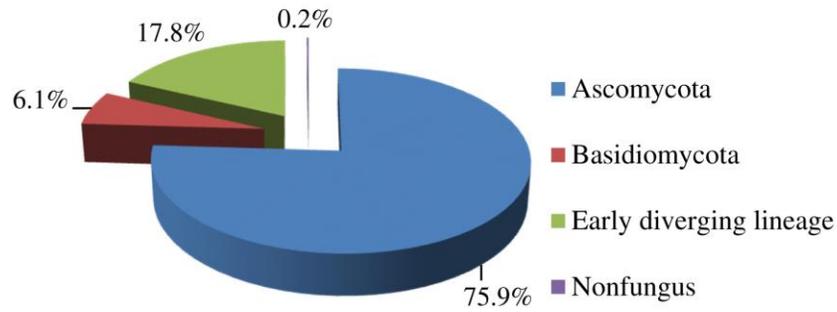

A) Culture-Dependent All Isolates

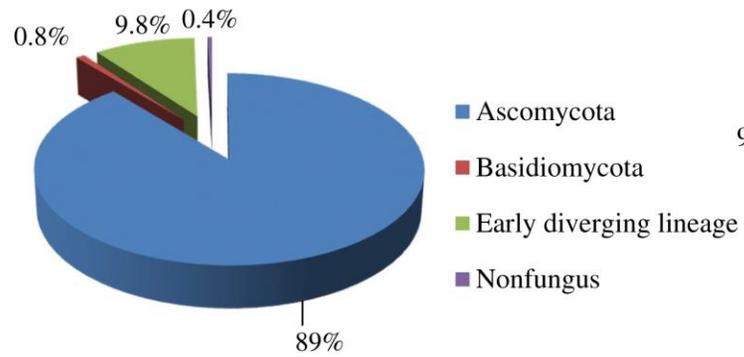

B) Culture-Dependent ITS Isolates

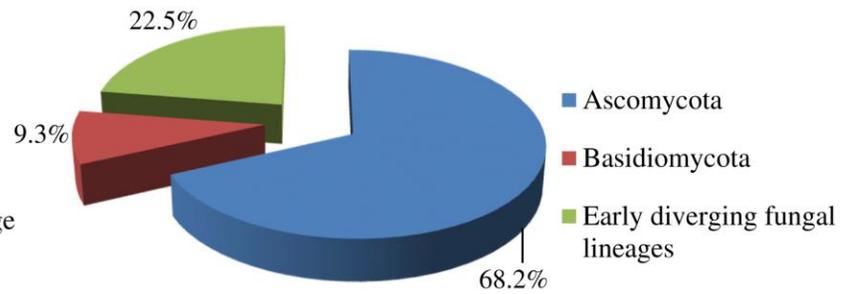

C) Culture-Dependent ITS2 Isolates

**Figure 2.**

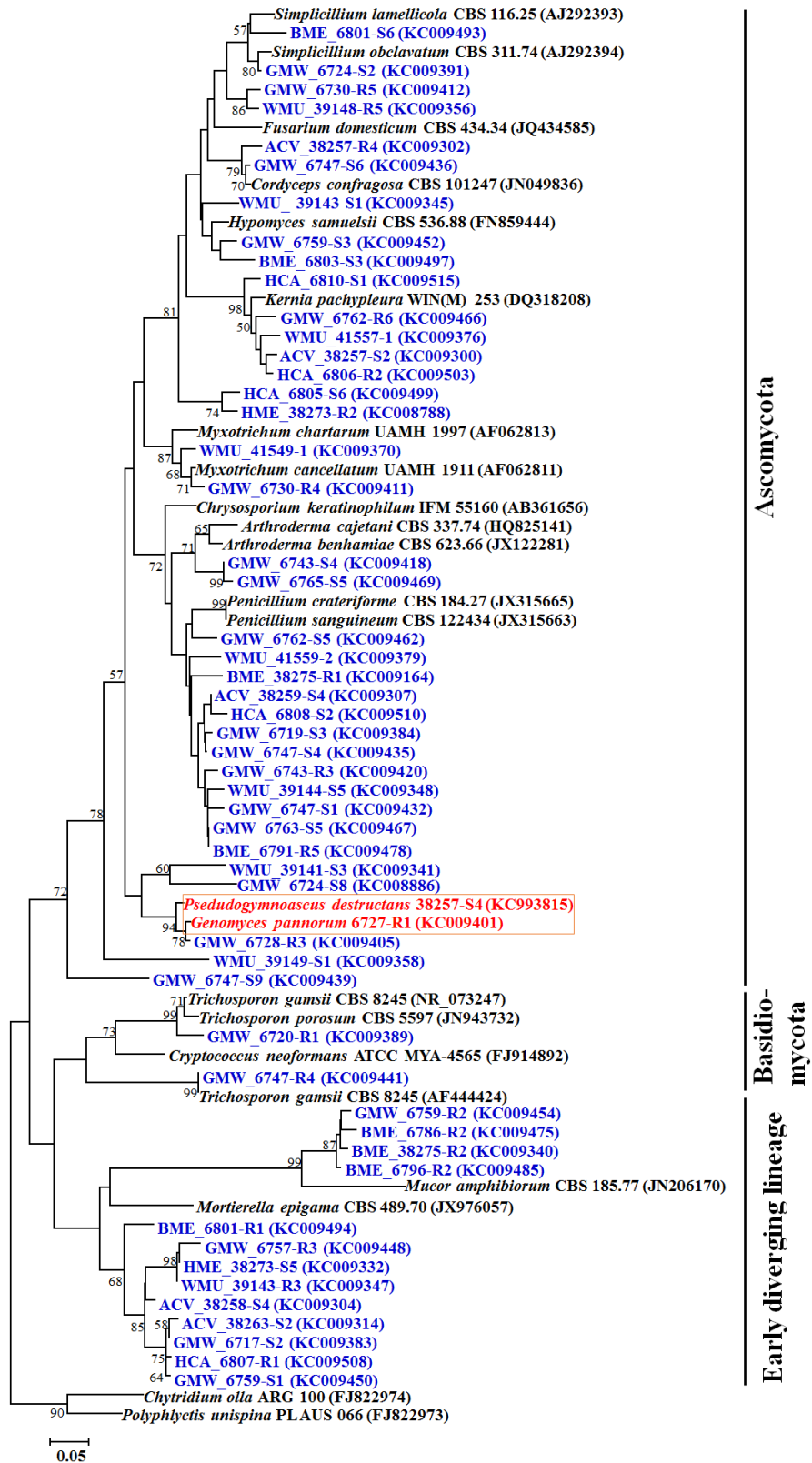

**Figure 3.**

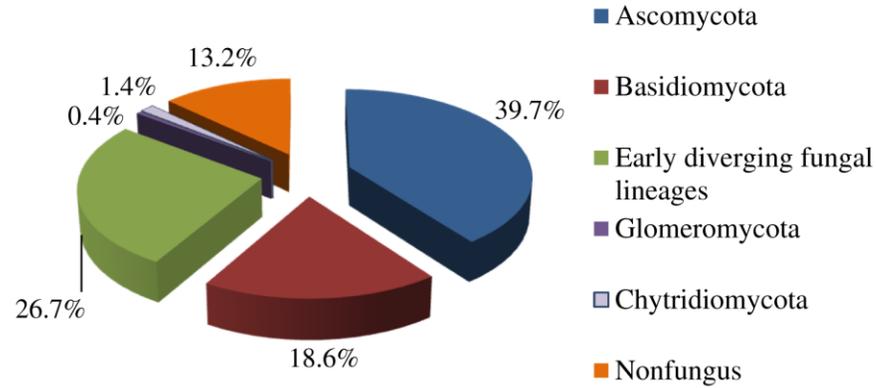

A) Culture-Independent All Clones

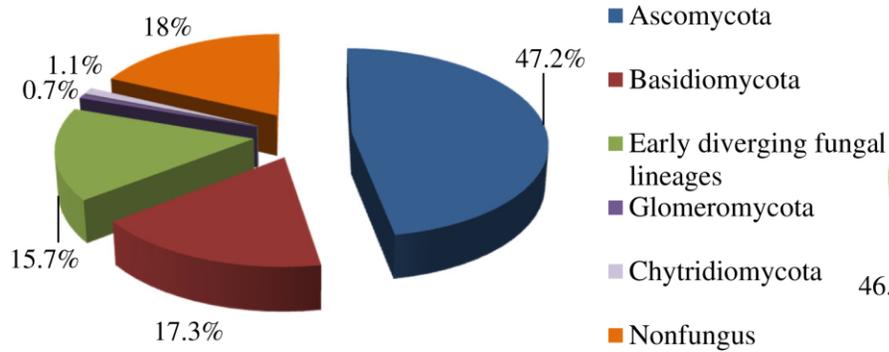

B) Culture-Independent LSU Clones

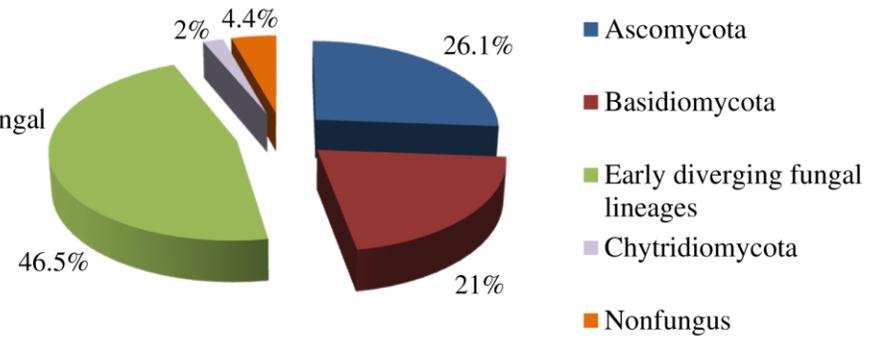

C) Culture-Independent ITS Clones

**Figure 4.**

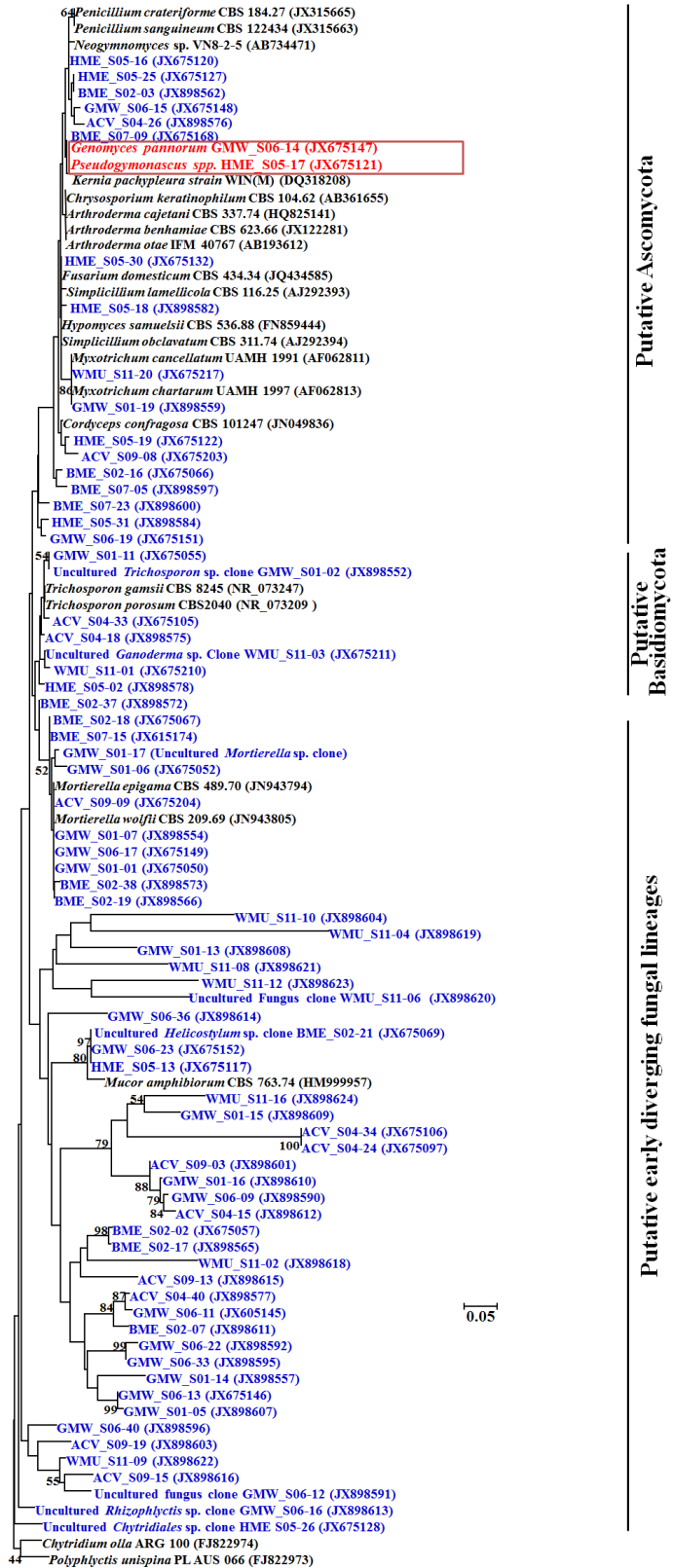

**Figure 5.**

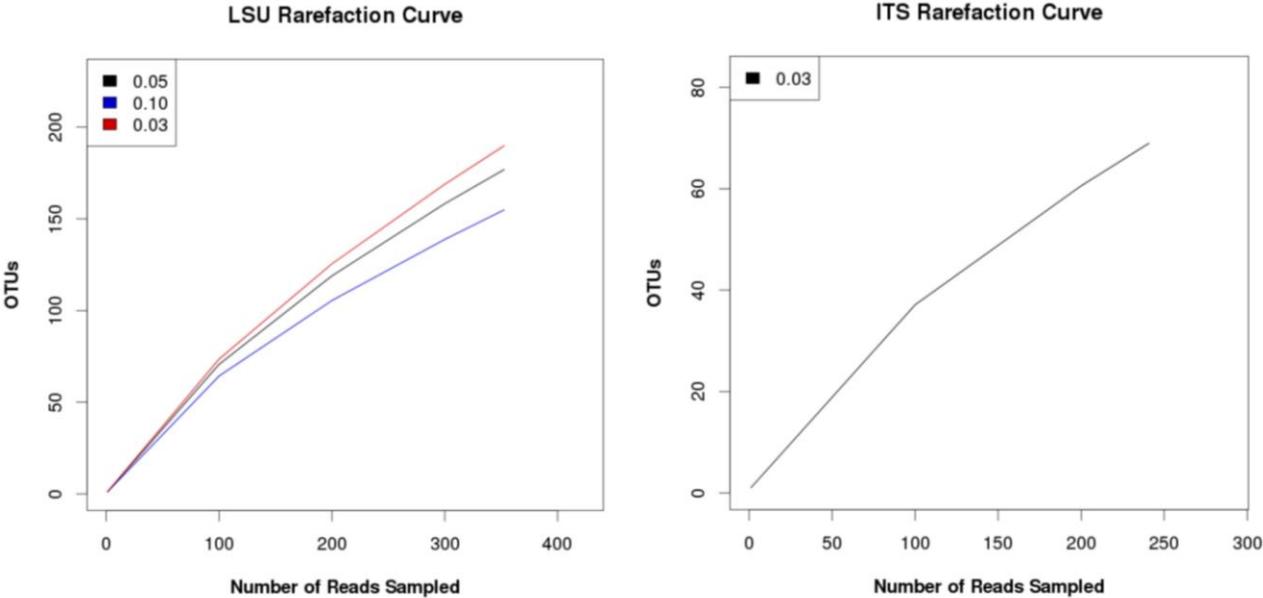

**Figure 6.**

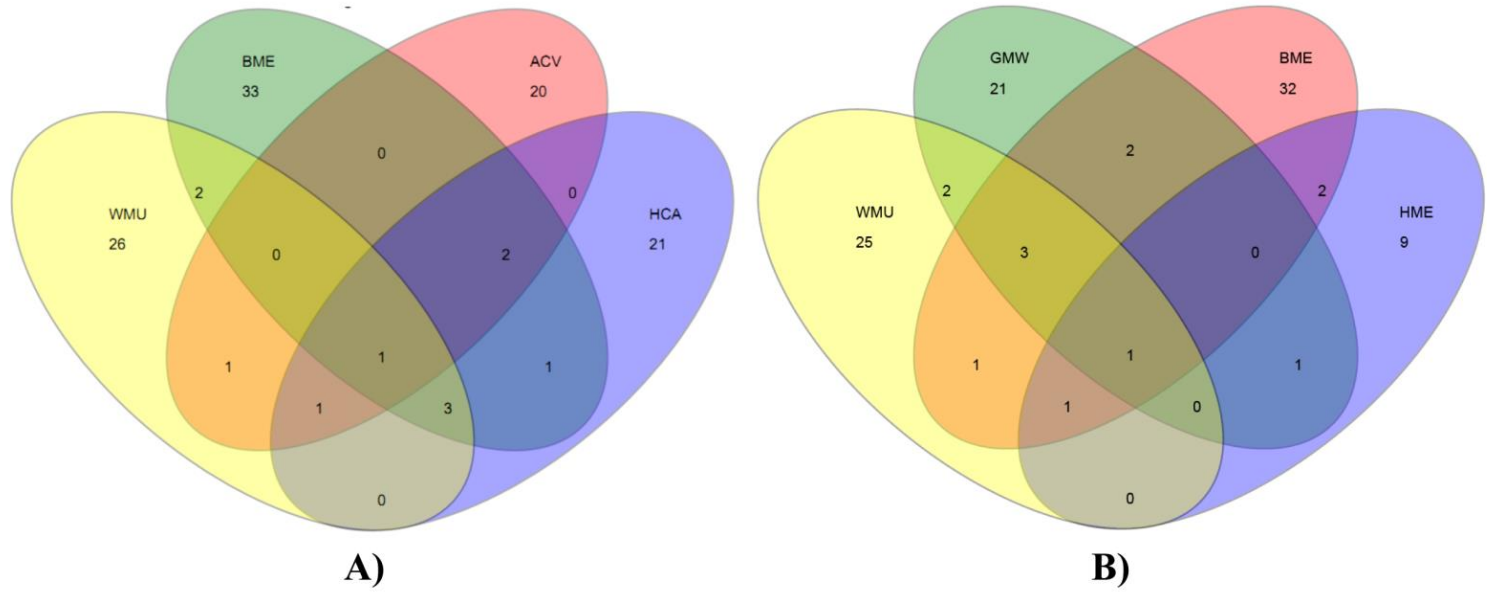

# Supplementary Data

Table S1. Bat hibernacula in the eastern United States surveyed for *Pd* by rt-PCR and CD methods

| NO. | Source | Collection | Sample Type | *Pd* by real time PCR | *Pd* by culture |
|---|---|---|---|---|---|
| 38249 | Aeolus Cave #1, E. Dorset, VT | Aug. 2010 | Decomposed bat remains | +* | -** |
| 38250 | Aeolus Cave #2, E. Dorset, VT | Aug. 2010 | bat remains and debris | + | - |
| 38251 | Aeolus Cave #3, E. Dorset, VT | Aug. 2010 | decomposed bat remains | - | - |
| 38252 | Aeolus Cave #4, E. Dorset, VT | Aug. 2010 | decomposed bat remains | + | - |
| 38253 | Aeolus Cave #5, E. Dorset, VT | Aug. 2010 | decomposed bat remains | + | - |
| 38254 | Aeolus Cave #6, E. Dorset, VT | Aug. 2010 | decomposed bat remains | + | - |
| 38255 | Aeolus Cave #7, E. Dorset, VT | Aug. 2010 | decomposed bat remains | + | - |
| 38257 | Aeolus Cave #8, E. Dorset, VT | Aug. 2010 | sediment under bat remains | + | + |
| 38258 | Aeolus Cave #9, E. Dorset, VT | Aug. 2010 | surface swab | + | - |
| 38259 | Aeolus Cave #10, E. Dorset, VT | Aug. 2010 | surface swab | + | - |
| 38261 | Aeolus Cave #11, E. Dorset, VT | Aug. 2010 | surface swab | - | - |
| 38263 | Aeolus Cave #12, E. Dorset, VT | Aug. 2010 | surface swab | + | - |
| 38264 | Aeolus Cave #13, E. Dorset, VT | Aug. 2010 | decomposed bat remains | + | - |
| 38266 | Aeolus Cave #14, E. Dorset, VT | Aug. 2010 | sediment under bat remains | + | - |
| 38268 | Aeolus Cave #15, E. Dorset, VT | Aug. 2010 | surface swab | + | - |
| 38270 | Hitchcock Mine #1, Paradox, Essex | Jan. 2010 | surf sediment and bat feces | + | - |
| 38273 | Hitchcock Mine #2, Paradox, Essex, NY | Jan. 2010 | surface sediment | + | - |
| 38274 | Barton Mine #1, Mineville, Essex, NY | Feb. 2010 | surface sediment | - | - |
| 38275 | Barton Mine #2, | Feb. 2010 | surface sediment | + | |
| | Mineville, Essex, NY | | | | |
| 39141 | Williams Preserve Mine, Kinston, Ulster, NY | Nov. 2010 | air | ND | + |
| 39142 | Williams Preserve Mine, Kinston, Ulster | Nov. 2010 | air | ND | - |
| 39143 | Williams Preserve Mine, Kinston, Ulster, NY | Nov. 2010 | air | ND | - |
| 39144 | Williams Preserve Mine, Kinston, Ulster, NY | Nov. 2010 | air | ND | - |
| 39145 | Williams Preserve Mine, Kinston, Ulster, NY | Nov. 2010 | swab | - | + |
| 39147 | Williams Preserve Mine, Kinston, Ulster, NY | Nov. 2010 | swab | - | + |
| 39148 | Williams Preserve Mine, Kinston, Ulster, NY | Nov. 2010 | swab | - | - |
| 39149 | Williams Preserve Mine, Kinston, Ulster, NY | Nov. 2010 | swab | - | - |
| 39150 | Williams Preserve Mine, Kinston, Ulster, NY | Nov. 2010 | soil | - | - |
| 39151 | Williams Preserve Mine, Kinston, Ulster, NY | Nov. 2010 | soil | - | - |
| 41544 | Williams Preserve Mine WP#1, Ulster, NY | Nov. 2010 | swab | - | + |
| 41546 | Williams Preserve Mine WP#2, Ulster, NY | Nov. 2010 | swab | - | - |
| 41547 | Williams Preserve Mine WP3A, Ulster, NY | Nov. 2010 | swab | + | + |
| 41548 | Williams Preserve Mine WP3B, Ulster, NY | Nov. 2010 | swab | + | + |
| 41549 | Williams Preserve Mine WP#4, Ulster, NY | Nov. 2010 | swab | - | - |
| 41551 | Williams Preserve | Nov. 2010 | swab | - | + |
| | Mine WP#6, Ulster, NY | | | | |
| 41552 | Williams Preserve Mine WP#7, Ulster, NY | Nov. 2010 | swab | - | - |
| 41553 | Williams Preserve Mine WP#8, Ulster, NY | Nov. 2010 | swab | + | - |
| 41554 | Williams Preserve Mine WP#9, Ulster, NY | Nov. 2010 | swab | - | - |
| 41555 | Williams Preserve Mine WP#3S, Ulster, NY | Nov. 2010 | sediment | - | - |
| 41556 | Williams Preserve Mine WP#4S, Ulster, NY | Nov. 2010 | sediment | - | + |
| 41557 | Williams Preserve Mine WP#5S, Ulster, NY | Nov. 2010 | sediment | - | - |
| 41558 | Williams Preserve Mine WP#6S, Ulster, NY | Nov. 2010 | sediment | - | + |
| 41559 | Williams Preserve Mine WP#8S, Ulster, NY | Nov. 2010 | swab | - | - |
| 41560 | Williams Preserve Mine WP#9S, Ulster, NY | Nov. 2010 | sediment | - | - |
| 6717 | Graphite Mine #1, Hague, Warren, NY | Jan. 2011 | swab | + | - |
| 6719 | Graphite Mine #3, Hague, Warren, NY | Jan. 2011 | swab | - | - |
| 6720 | Graphite Mine #5, Hague, Warren, NY | Jan. 2011 | swab | + | - |
| 6724 | Graphite Mine #6, Hague, Warren, NY | Jan. 2011 | sediment | + | + |
| 6727 | Graphite Mine #8, Hague, Warren, NY | Jan. 2011 | swab | + | - |
| 6728 | Graphite Mine #10, Hague, Warren, NY | Jan. 2011 | swab | - | - |
| 6730 | Graphite Mine #12, Hague, Warren, NY | Jan. 2011 | swab | - | - |
| 6732 | Graphite Mine #15, Hague, Warren, NY | Jan. 2011 | sediment | - | - |
| 6733 | Graphite Mine #17, Hague, Warren, NY | Jan. 2011 | sed/swab? | + | - |
| 6734 | Graphite Mine #19, Hague, Warren, NY | Jan. 2011 | swab | + | + |
| 6743 | Graphite Mine #20, Hague, Warren, NY | Jan. 2011 | sediment | + | - |
| 6745 | Graphite Mine #23, Hague, Warren, NY | Jan. 2011 | swab | + | + |
| 6747 | Graphite Mine #25, Hague, Warren, NY | Jan. 2011 | swab | - | - |
| 6757 | Graphite Mine #26, Hague, Warren, NY | Jan. 2011 | swab | - | + |
| 6759 | Graphite Mine #27, Hague, Warren, NY | Jan. 2011 | sediment | - | - |
| 6762 | Graphite Mine #29, Hague, Warren, NY | Jan. 2011 | sediment | - | - |
| 6763 | Graphite Mine #30, Hague, Warren, NY | Jan. 2011 | swab | + | + |
| 6765 | Graphite Mine #32, Hague, Warren, NY | Jan. 2011 | sediment | + | + |
| 6767 | Graphite Mine #33, Hague, Warren, NY | Jan. 2011 | sediment | - | - |
| 6777 | Barton Mine #1, Moriah, Essex, NY | Feb. 2011 | sediment | - | - |
| 6785 | Barton Mine #3, Moriah, Essex, NY | Feb. 2011 | sediment | + | - |
| 6786 | Barton Mine #5, Moriah, Essex, NY | Feb. 2011 | sediment | - | - |
| 6789 | Barton Mine #6, Moriah, Essex, NY | Feb. 2011 | swab | + | - |
| 6790 | Barton Mine #7, Moriah, Essex, NY | Feb. 2011 | swab | - | - |
| 6791 | Barton Mine #10, Moriah, Essex, NY | Feb. 2011 | sediment | - | - |
| 6792 | Barton Mine #11, Moriah, Essex, NY | Feb. 2011 | sediment | + | - |
| 6793 | Barton Mine #12, Moriah, Essex, NY | Feb. 2011 | swab | + | - |
| 6794 | Barton Mine #14, Moriah, Essex, NY | Feb. 2011 | swab | - | - |
| 6795 | Barton Mine #18, Moriah, Essex, NY | Feb. 2011 | swab | + | - |
| 6796 | Barton Mine #22, Moriah, Essex, NY | Feb. 2011 | swab | + | - |
| 6797 | Barton Mine #23, Moriah, Essex, NY | Feb. 2011 | sediment | - | - |
| 6798 | Barton Mine #25, Moriah, Essex, NY | Feb. 2011 | sediment | + | - |
| 6799 | Barton Mine #27, | Feb. 2011 | swab | - | - |
| | Moriah, Essex, NY | | | | |
| 6801 | Barton Mine #28, Moriah, Essex, NY | Feb. 2011 | swab | + | - |
| 6802 | Barton Mine #31, Moriah, Essex, NY | Feb. 2011 | sediment | - | - |
| 6803 | Barton Mine #33, Moriah, Essex, NY | Feb. 2011 | sediment | - | - |
| 6804 | Barton Mine #39, Moriah, Essex, NY | Feb. 2011 | sediment | - | - |
| 6805 | Hailes Cave #1, Guilderland, Albany, NY | Feb. 2011 | sediment | - | - |
| 6806 | Hailes Cave #2, Guilderland, Albany, NY | Feb. 2011 | swab | - | - |
| 6807 | Hailes Cave #4, Guilderland, Albany, NY | Feb. 2011 | swab | + | - |
| 6808 | Hailes Cave #5, Guilderland, Albany, NY | Feb. 2011 | swab | - | - |
| 6810 | Hailes Cave #6, Guilderland, Albany, NY | Feb. 2011 | swab | + | - |
| 6811 | Hailes Cave #9, Guilderland, Albany, NY | Feb. 2011 | swab | + | - |
| 6812 | Hailes Cave #10, Guilderland, Albany, NY | Feb. 2011 | swab | + | - |
| 6814 | Hailes Cave #14, Guilderland, Albany | Feb. 2011 | swab | - | - |
| 6817 | Hailes Cave #22, Guilderland, Albany, NY | Feb. 2011 | Dry feces/sediment (racoon?) | - | - |

Note: "?", indicates the type of sample was not clear. * denotes positive (+) or negative (-) for *Pd* gDNA; ** denotes positive (+) or negative (-) for *Pd* culture; CD, culture-dependent; rt-PCR, real-time PCR; ND, not done

Table S2. Environmental samples used for library construction

| Source | Sample type (Sample no.) | LSU clone | ITS clone | *Pd* real time PCR |
|---|---|---|---|---|
| Graphite mine Warren, NY (GMW) | Sediment (6724, 6743) | GMW_D01 | GMW_S01 | + |
| | Swab (6734, 6745) | GMW_D06 | GMW_S06 | + |
| Barton mine, Essex, NY (BME) | Sediment (6785, 6792) | BME_D02 | BME_S02 | + |
| | Swab (6789, 6801) | BME_D07 | BME_S07 | + |
| | Swab (38274, 38275) | BME_D010 | BME_S10 | -/+ |
| Hailes cave, Albany, NY (HCA) | Sediment (6805, 6817) | HCA_D03 | HCA_S03 | - |
| | Swab (6807, 6811) | HCA_D08 | HCA_S08 | + |
| Aeolus cave, Bennington, VT (ACV) | Sediment (38257, 38266) | ACV_D04 | ACV_S04 | + |
| | Swab (38258, 38263) | ACV_D09 | ACV_S09 | + |
| Williams mine, Ulster, NY (WMU) | Sediment (39150, 39151) | WMU_D011 | WMU_S11 | - |
| | Swab (39145, 41549) | WMU_D012 | WMU_S12 | -/+ |
| Hitchcock mine, Essex, NY (HME) | Sediment (38270, 38273) | HME_D05 | HME_S05 | + |

Note: The sample number in parenthesis under the sample type column are same source as described in table S1; positive (+) or negative (-) for presence or absence of *Pd* gDNA

Table S3. GenBank accession numbers of ribosomal gene sequences by CD and CI methods

| | Target | Parameters | Numbers | Accession numbers |
|---|---|---|---|---|
| **Cultures** | ITS | ITS2 | 399 | KC008730-KC009128 |
| | | ITS | 237 | KC009286-KC009522 |
| | LSU | LSU | 39 | KC009247-KC009285 |
| **Culture-independent clones** | LSU | Fungal confirmed sequences | 330 | JX534602-JX534931 |
| | | Fungal sequences | 40 | JX898636-JX898675 |
| | | Non fungal sequences | 62 | JX545249-JX545310 |
| | | Chimera sequences non fungal | 19 | JX898533-JX898551 |
| | ITS | 100 % coverage sequences | 168 | JX675050-JX675217 |
| | | 50-99% coverage sequences | 55 | JX898552-JX898606 |
| | | <50% coverage sequences | 18 | JX898607-JX898624 |
| | | Non fungal sequences | 11 | JX898625-JX898635 |

Table S4. Details of ITS2 sequences of fungal isolates recovered by CD method

| Sum (%)[a] | OTU[b] | Accession no. | Best BLAST hit Taxon | Phylum | Class | Score[c] | Acc. no.[d] | % Similiar[e] |
|---|---|---|---|---|---|---|---|---|
| 0.42 | 6801-S7 | KC009061 | *Mucor hiemalis* | EDFL | Zygomycetes | 366 | JX845571 | 100 |
| 0.42 | *6801-S6 | KC009493 | *Verticillium leptobactrum* | Ascomycota | Deuteromycetes | 592 | EF641874 | 98 |
| 0.42 | *6801-R1 | KC009494 | *Mortierella polycephala* | EDFL | Zygomycetes | 711 | HQ630334 | 98 |
| 0.42 | 6803-S3 | KC009497 | *Verticillium* sp. | Ascomycota | Deuteromycetes | 443 | DQ888742 | 99 |
| 0.42 | 6805-S6 | KC009499 | *Podospora* sp. | Ascomycota | Sordariomycetes | 500 | JN689974 | 99 |
| 0.42 | 6806-R2 | KC009503 | *Kernia* sp. | Ascomycota | Sordariomycetes | 607 | FJ946487 | 98 |
| 0.42 | 6807-R1 | KC009508 | *Mortierella* sp. | EDFL | Zygomycetes | 731 | HQ630307 | 100 |
| 0.42 | 39143-R3 | KC009347 | *Mortierella* sp. | EDFL | Zygomycetes | 737 | EU240130 | 99 |
| 0.42 | *39144-S5 | KC009348 | *Penicillium brevicompactum* | Ascomycota | Eurotiomycetes | 625 | AY373898 | 99 |
| 0.42 | 39148-R5 | KC009356 | *Cosmospora* sp. | Ascomycota | Sordariomycetes | 614 | JN995627 | 99 |
| 0.42 | 39149-S1 | KC009358 | *Podospora* sp. | Ascomycota | Sordariomycetes | 601 | HQ647346 | 99 |
| 0.42 | 6808-S2 | KC009510 | *Aspergillus asperescens* | Ascomycota | Eurotiomycetes | 61 | EF652475 | 100 |
| 0.42 | *38273-S5 | KC009332 | *Mortierella parvispora* | EDFL | Zygomycetes | 737 | EU484279 | 99 |
| 0.42 | *38273-R2 | KC009334 | *Chaetomium crispatum* | Ascomycota | Sordariomycetes | 612 | HM365267 | 100 |
| 0.42 | 38275-R1 | KC009339 | *Gymnoascus* sp. | Ascomycota | Eurotiomycetes | 352 | AB361643 | 94 |
| 0.42 | 6759-S3 | KC009452 | *Hypomyces aurantius* | Ascomycota | Sordariomycetes | 482 | AB591044 | 99 |
| 0.42 | *6759-R2 | KC009454 | *Mucor flavus* | EDFL | Zygomycetes | 637 | EU484282 | 98 |
| 0.42 | 6762-S5 | KC009462 | *Chrysosporium* sp. | Ascomycota | Eurotiomycetes | 556 | AM949568 | 95 |
| 0.42 | 6762-R6 | KC009466 | *Wardomyces humicola* | Ascomycota | Sordariomycetes | 590 | AM774157 | 97 |
| 0.42 | *6763-S5 | KC009467 | *Penicillium swiecickii* | Ascomycota | Eurotiomycetes | 601 | GU441580 | 99 |
| 0.42 | 6765-S5 | KC009469 | *Arthroderma* sp. | Ascomycota | Euascomycetes | 518 | AJ877216 | 92 |
| 0.42 | 6786-R2 | KC009475 | *Mucoromycote* sp. | EDFL | Zygomycetes | 461 | EF555501 | 84 |
| 0.42 | 39141-R2 | KC009343 | *Penicillium polonicum* | Ascomycota | Eurotiomycetes | 540 | JN368451 | 100 |
| 0.42 | 41554-4 | KC009372 | *Mortierella turficola* | EDFL | Zygomycetes | 614 | HQ630350 | 97 |
| 0.42 | 6724-S2 | KC009391 | *Simplicillium* sp. | Ascomycota | Sordariomycetes | 590 | AB604004 | 97 |
| 0.42 | *6727-R1 | KC009401 | *Geomyces pannorum* | Ascomycota | Leotiomycetes | 605 | HQ115661 | 100 |
| 0.42 | 6747-S9 | KC009439 | *Debaryomyces hansenii* | Ascomycota | Saccharomycetes | 684 | JQ912667 | 100 |
| 0.42 | 6810-R7 | KC009518 | *Kernia* sp. | Ascomycota | Sordariomycetes | 515 | DQ318208 | 97 |
| 0.42 | 6743-R3 | KC009420 | *Penicillium raphiae* | Ascomycota | Eurotiomycetes | 639 | JN617673 | 100 |
| 0.42 | 6747-S1 | KC009432 | *Penicillium virgatum* | Ascomycota | Eurotiomycetes | 594 | JF439503 | 98 |
| 0.42 | 6747-S4 | KC009435 | *Penicillium angulare* | Ascomycota | Eurotiomycetes | 583 | AY313613 | 97 |
| 0.42 | 6757-R3 | KC009448 | *Mortierella* sp. | EDFL | Zygomycetes | 493 | JQ670951 | 94 |
| 0.42 | 6717-S2 | KC009383 | *Mortierella* sp. | EDFL | Zygomycetes | 682 | AY842393 | 95 |
| 0.84 | 6791-R5 | KC009478 | *Penicillium sanguifluum* | Ascomycota | Eurotiomycetes | 634 | JN617711 | 99 |
| 0.84 | 38257-R4 | KC009302 | *Cordyceps militaris* | Ascomycota | Sordariomycetes | 585 | AF122036 | 97 |
| 0.84 | 38263-S2 | KC009314 | *Verticillium* sp. | Ascomycota | Deuteromycetes | 751 | FJ025166 | 98 |
| 0.84 | 39141-S3 | KC009341 | *Cladosporium cladosporioides* | Ascomycota | Dothideomycetes | 524 | JX077073 | 100 |
| 0.84 | *39143-S1 | KC009345 | *Hypocrea pachybasioides* | Ascomycota | Sordariomycetes | 619 | JX406549 | 100 |
| 0.84 | *38258-S4 | KC009304 | *Mortierella alpina* | EDFL | Zygomycetes | 755 | AB476415 | 100 |
| 0.84 | 6747-S6 | KC009436 | *Isaria farinosa* | Ascomycota | Sordariomycetes | 630 | HQ115724 | 100 |
| 0.84 | *6747-R4 | KC009441 | *Guehomyces pullulans* | Ascomycota | Tremellomycetes | 682 | AF444417 | 99 |
| 0.84 | 41559-2 | KC009379 | *Arachniotus aurantiacus* | Ascomycota | Eurotiomycetes | 610 | HM991267 | 98 |
| 1.26 | 6719-S3 | KC009384 | *Penicillium thomii* | Ascomycota | Eurotiomycetes | 627 | JN585937 | 100 |
| 1.26 | 6759-S1 | KC009450 | *Mortierella* sp. | EDFL | Zygomycetes | 697 | AY157495 | 95 |
| 1.26 | 38257-S2 | KC009300 | *Wardomyces inflatus* | Ascomycota | Sordariomycetes | 666 | HQ914934 | 99 |
| 1.68 | 6730-R4 | KC009411 | *Myxotrichum* sp. | Ascomycota | Leotiomycetes | 542 | AF062815 | 97 |
| 2.10 | *38275-R2 | KC009340 | *Helicostylum pulchrum* | EDFL | Zygomycetes | 704 | AB614353 | 99 |
| 2.10 | 6796-R2 | KC009485 | *Helicostylum* sp. | EDFL | Zygomycetes | 569 | AB614353 | 93 |
| 2.10 | 41557-1 | KC009376 | *Doratomyces stemonitis* | Ascomycota | Sordariomycetes | 518 | JN104543 | 100 |
| 2.10 | 6728-R3 | KC009405 | *Polypaecilum botryoides* | Ascomycota | Eurotiomycetes | 494 | CBS 176.44 | 100 |
| 2.52 | *6720-R1 | KC009389 | *Trichosporon dulcitum* | Ascomycota | Tremellomycetes | 763 | NR_073248 | 100 |
| 2.52 | 6730-R | KC009412 | *Fusarium merismoides* | Ascomycota | Sordariomycetes | 535 | EU860057 | 100 |
| 6.30 | 6743-S4 | KC009418 | *Arthroderma* sp. | Ascomycota | Euascomycetes | 480 | JN104536 | 94 |
| 7.98 | 6810-S1 | KC009515 | *Kernia* sp. | Ascomycota | Sordariomycetes | 502 | DQ318208 | 97 |
| 7.98 | *41549-1 | KC009370 | *Oidiodendron truncatum* | Ascomycota | Leotiomycetes | 592 | FJ914713 | 100 |
| 11.8 | *38270-7 | KC009329 | *Geomyces pannorum* | Ascomycota | Leotiomycetes | 489 | AB517942 | 99 |
| 25.6 | 38259-S4 | KC009307 | *Penicillium polonicum* | Ascomycota | Eurotiomycetes | 634 | JQ082508 | 100 |

[a] Relative abundance for the combined libraries, which was used to sort the entries.
[b] OTUs were characterized by Mothur program [1], the OTU is ≥97% similar to a fungal isolate.
[c] BLASTN [2] score value.
[d] Accession number of the closest database match.
[e] Level of similarity for pairwise alignments with the closest match, using the Martinez-Needleman-Wunsch algorithm [2].
*Common OTUs identified from ITS2, ITS, and LSU sequences

Table S5. Details of ITS sequences of fungal isolates recovered by CD method

| Sum (%)[a] | OTU[b] | Accession no. | Best BLAST hit | | | Score[c] | Acc. No. | % Similiar[e] |
|---|---|---|---|---|---|---|---|---|
| | | | Taxon | Phylum | Class | | | |
| 0.25 | 38249-2 | KC008730 | *Phoma sclerotioides* | Ascomycota | Dothideomycetes | 726 | FJ179161 | 99 |
| 0.25 | 38249-8 | KC008734 | *Mortierella stylospora* | EDFL | Zygomycetes | 507 | JX976086 | 88 |
| 0.25 | 38251-R1 | KC008746 | *Arthroderma silverae* | Ascomycota | Eurotiomycetes | 608 | AJ877216 | 89 |
| 0.25 | 38254-S4 | KC008757 | *Mortierella fimbricystis* | EDFL | Zygomycetes | 904 | JX976069 | 95 |
| 025 | 38253-R3 | KC008754 | *Neonectria ramulariae* | Ascomycota | Sordariomycetes | 854 | AY781232 | 98 |
| 0.25 | 39147-S1 | KC008820 | *Pleosporales sp.* | Ascomycota | Dothideomycetes | 767 | HQ713762 | 99 |
| 0.25 | 39142-S4 | KC008804 | *Dothiorella gregaria* | Ascomycota | Dothideomycetes | 793 | AB470899 | 100 |
| 0.25 | 6777-S7 | KC008970 | *Aphanocladium album* | Ascomycota | Sordariomycetes | 545 | KC291734 | 90 |
| 0.25 | 6805-S1 | KC009082 | *Bionectria ochroleuca* | Ascomycota | Sordariomycetes | 817 | JN002172 | 99 |
| 0.25 | 41559-3 | KC008871 | *Simplicillium lanosoniveum* | Ascomycota | Sordariomycetes | 701 | KC172069 | 92 |
| 0.25 | 6717-S1 | KC008872 | *Verticillium leptobactrum* | Ascomycota | Sordariomycetes | 881 | EF641870 | 99 |
| 0.25 | 6811-S2 | KC009105 | *Beauveria bassiana* | Ascomycota | Sordariomycetes | 872 | KC753398 | 100 |
| 0.25 | 6806-S6 | KC009090 | *Podospora sp.* | Ascomycota | Sordariomycetes | 647 | AY999127 | 93 |
| 0.25 | 39144-S6 | KC008813 | *Leotiomycetes sp.* | Ascomycota | Leotiomycetes | 641 | AB190398 | 99 |
| 0.25 | 39144-S2 | KC008810 | *Candida pseudoglaebosa* | Ascomycota | Saccharomycetales | 915 | NR_111588 | 99 |
| 0.25 | *6811-S1 | KC009104 | *Mortierella alpina* | EDFL | Zygomycetes | 1013 | KF313129 | 100 |
| 0.25 | 41554-2 | KC008862 | *Mortierellales sp.* | EDFL | Zygomycetes | 1033 | DQ865089 | 100 |
| 0.25 | 6806-R5 | KC009093 | *Mortierellaceae sp.* | EDFL | Zygomycetes | 939 | JX975843 | 99 |

| 0.25 | 6803-R5 | KC009072 | *Mortierella histoplasmatoides* | EDFL | Zygomycetes | 939 | HQ630309 | 99 |
| --- | --- | --- | --- | --- | --- | --- | --- | --- |
| 0.25 | 39141-S5 | KC008799 | *Trichophyton terrestre* | Ascomycota | Eurotiomycetes | 1016 | EF568097 | 99 |
| 0.25 | *6797-R1 | KC009053 | *Mortierella sarnyensis* | EDFL | Zygomycetes | 902 | JF311973 | 97 |
| 0.25 | 38268-S2 | KC008781 | *Mortierella sp.* | EDFL | Zygomycetes | 490 | JX975922 | 86 |
| 0.25 | *6786-R3 | KC008990 | *Mortierella polycephala* | EDFL | Zygomycetes | 1029 | JX975900 | 100 |
| 0.25 | 41547-1 | KC008844 | *Cryptococcus fragicola* | Basidiomycota | Tremellomycetes | 730 | AB035588 | 100 |
| 0.25 | 6794-R4 | KC009032 | *Holtermanniella watticus* | Basidiomycota | Tremellomycetes | 870 | JQ857031 | 100 |
| 0.25 | 39142-S2 | KC008802 | *Mucor luteus* | EDFL | Zygomycetes | 929 | NR_103614 | 97 |
| 0.25 | 41549-4 | KC008849 | *Mucor flavus* | EDFL | Zygomycetes | 1026 | JN206049 | 98 |
| 0.25 | *6805-R2 | KC009087 | *Mucor flavus* | EDFL | Zygomycetes | 1110 | JN206061 | 99 |
| 0.25 | 41544-1 | KC008832 | *Mortierella fimbricystis* | EDFL | Zygomycetes | 785 | JX976069 | 93 |
| 0.25 | 6815-R3 | KC009121 | *Umbelopsis isabellina* | EDFL | Zygomycetes | 996 | JF303862 | 99 |
| 0.25 | 41548-3 | KC008848 | *Leuconeurospora pulcherrima* | EDFL | Zygomycetes | 900 | KF049206 | 99 |
| 0.25 | 6791-R1 | KC009010 | *Penicillium griseofulvum* | Ascomycota | Eurotiomycetes | 867 | DQ339549 | 99 |
| 0.25 | *6717-R1 | KC008874 | *Mortierella hyalina* | EDFL | Zygomycetes | 822 | JX975983 | 97 |
| 0.25 | *6807-S2 | KC009095 | *Penicillium quercetorum* | Ascomycota | Eurotiomycetes | 819 | AY443471 | 98 |
| 0.25 | 39143-R1 | KC008808 | *Botryotinia fuckeliana* | Ascomycota | Leotiomycetes | 815 | KF802809 | 99 |
| 0.25 | 6763-S6 | KC008950 | *Oidiodendron tenuissimum* | Ascomycota | Leotiomycetes | 734 | NR_111036 | 95 |
| 0.25 | 6806-S2 | KC009089 | *Penicillium glabrum* | Ascomycota | Leotiomycetes | 883 | JX140803 | 99 |
| 0.25 | 6789-S4 | KC008996 | *Penicillium miczynskii* | Ascomycota | Leotiomycetes | 915 | JN617670 | 100 |
| 0.25 | 6808-S5 | KC009098 | *Oidiodendron sp.* | Ascomycota | Leotiomycetes | 695 | NR_111036 | 94 |
| 0.25 | 6765-S8 | KC008957 | *Penicillium tularense* | Ascomycota | Leotiomycetes | 806 | KC427195 | 98 |
| 0.25 | 6804-R2 | KC009078 | *Hypocrea viridescens* | Ascomycota | Sordariomycetes | 893 | KJ482546 | 100 |
| 0.25 | 6719-S4 | KC008877 | *Zalerion varium* | Ascomycota | Sordariomycetes | 832 | KF156329 | 100 |
| 0.25 | 6743-R2 | KC008924 | *Neobulgaria sp.* | Ascomycota | Leotiomycetes | 845 | HM051080 | 98 |
| 0.25 | 41558-2 | KC008869 | *Doratomyces sp.* | Ascomycota | Sordariomycetes | 898 | FJ914706 | 99 |
| 0.25 | 6727-S1 | KC008889 | *Chrysosporium vallenarense* | Ascomycota | Eurotiomycetes | 527 | NR_077139 | 85 |
| 0.5 | 6785-S7 | KC008978 | *Auxarthron umbrinum* | Ascomycota | Eurotiomycetes | 749 | NR_111136 | 96 |
| 0.5 | 38250-S5 | KC008742 | *Geomyces pannorum* | Ascomycota | Leotiomycetes | 924 | DQ494320 | 99 |
| 0.5 | *38273-R2 | KC008788 | *Chaetomium crispatum* | Ascomycota | Sordariomycetes | 811 | JQ864439 | 98 |
| 0.5 | 39144-S4 | KC008812 | *Penicillium vulpinum* | Ascomycota | Leotiomycetes | 564 | FJ004321 | 100 |
| 0.5 | 6785-S5 | KC008977 | *Myxotrichum sp.* | Ascomycota | Leotiomycetes | 708 | AF062815 | 94 |
| 0.5 | 6724-S1 | KC008883 | *Chrysosporium sp.* | Ascomycota | Eurotiomycetes | 603 | HF548536 | 87 |
| 0.5 | *38252-S1 | KC008748 | *Kernia pachypleura* | Ascomycota | Sordariomycetes | 725 | DQ318208 | 97 |
| 0.5 | *39141-S1 | KC008796 | *Hypocrea pachybasioides* | Ascomycota | Sordariomycetes | 941 | FJ860796 | 100 |
| 0.5 | 6817-S4 | KC009125 | *Mucor hiemalis* | EDFL | Zygomycetes | 913 | JX976246 | 97 |
| 0.75 | 6747-R2 | KC008936 | *Trichosporon shinodae* | Basidiomycota | Tremellomycetes | 730 | AB180201 | 99 |
| 0.75 | *6743-S5 | KC008923 | *Guehomyces pullulans* | Basidiomycota | Tremellomycetes | 896 | KC009015 | 99 |
| 0.75 | 39150-R1 | KC008833 | *Mortierella cf. gamsii* | EDFL | Zygomycetes | 976 | JX975892 | 99 |
| 0.75 | 39150-R2 | KC008834 | *Tetracladium sp.* | Ascomycota | Mitosporic Ascomycota | 826 | KC180672 | 99 |
| 0.75 | 41544-4 | KC008840 | *Mortierella dichotoma* | EDFL | Zygomycetes | 863 | JX975842 | 99 |
| 1.0 | 6719-S5 | KC008878 | *Mortierella hyalina* | EDFL | Zygomycetes | 1022 | JX975928 | 100 |
| 1.0 | 38264-S4 | KC008778 | *Kernia sp.* | Ascomycota | Sordariomycetes | 604 | DQ318208 | 97 |
| 1.25 | *39148-S1 | KC008723 | *Cladosporium cladosporioides* | Ascomycota | Dothideomycetes | 841 | KJ598781 | 100 |
| 1.25 | 38258-S2 | KC008767 | *Fusarium merismoides* | Ascomycota | Sordariomycetes | 861 | KC427027 | 100 |
| 1.5 | 6804-S1 | KC009075 | *Hypocrea viridescens* | Ascomycota | Sordariomycetes | 893 | KJ482546 | 100 |
| 1.5 | 38253-S4 | KC008752 | *Mortierella amoeboidea* | EDFL | Zygomycetes | 1042 | JX976068 | 100 |
| 1.5 | *6792-S4 | KC009014 | *Penicillium brevicompactum* | Ascomycota | Leotiomycetes | 905 | AY373899 | 100 |
| 1.5 | *38255-S2 | KC008761 | *Mucor hiemalis* | EDFL | Zygomycetes | 1003 | KF944455 | 100 |
| 1.5 | 41548-1 | KC008847 | *Cadophora fastigiata* | EDFL | Zygomycetes | 889 | EU484297 | 95 |
| 1.5 | 38261-R2 | KC008775 | *Mortierella gamsii* | EDFL | Zygomycetes | 926 | JX975984 | 98 |
| 1.5 | 38259-R2 | KC008769 | *Helicostylum elegans* | EDFL | Zygomycetes | 1029 | AB113014 | 99 |
| 1.75 | 6717-S3 | KC008873 | *Phaeothecoidea sp.* | Ascomycota | Dothideomycetes | 604 | KC460848 | 93 |

| 2.75 | *39147-R4 | KC008822 | *Helicostylum pulchrum* | EDFL | Zygomycetes | 876 | JN206053 | 99 |
|---|---|---|---|---|---|---|---|---|
| 2.75 | 38273-S4 | KC008787 | *Mortierella jenkinii* | EDFL | Zygomycetes | 965 | JX975849 | 99 |
| 3.0 | *38250-S1 | KC008739 | *Penicillium swiecickii* | Ascomycota | Leotiomycetes | 787 | AM236585 | 99 |
| 3.25 | 38250-S2 | KC008740 | *Mortierella sp.* | EDFL | Zygomycetes | 843 | JX976069 | 95 |
| 3.25 | 6785-S3 | KC008975 | *Debaryomyces maramus* | Ascomycota | Sordariomycetes | 989 | AJ586525 | 99 |
| 3.25 | 38254-R3 | KC008759 | *Geomyces sp.* | Ascomycota | Leotiomycetes | 658 | NR_111872 | 93 |
| 3.5 | *38252-R2 | KC008750 | *Oidiodendron truncatum* | Ascomycota | Leotiomycetes | 852 | FJ914713 | 99 |
| 7.5 | *38263-S2 | KC008776 | *Trichosporon dulcitum* | Basidiomycota | Tremellomycetes | 693 | NR_073248 | 100 |
| 13.25 | *38251-S3 | KC008745 | *Geomyces pannorum* | Ascomycota | Leotiomycetes | 833 | KC461530 | 100 |
| 21 | 38249-4 | KC008731 | *Penicillium griseofulvum* | Ascomycota | Leotiomycetes | 743 | JX091405 | 99 |

[a]Relative abundance for the combined libraries, which was used to sort the entries
[b]OTUs were characterized by Mothur program [1], the OTU is ≥97% similar to a fungal isolate
[c]BLASTN [2] score value
[d]Accession number of the closest database match
[e]Level of similarity for pairwise alignments with the closest match, using the Martinez-Needleman-Wunsch algorithm [2]
*Common OTUs identified from ITS2, ITS, and LSU sequences

Table S6. Details of LSU sequences of fungal isolates recovered by CD method

| Sum (%)[a] | OTU[b] | Accession no. | Taxon Coverage | Best BLAST hit Phylum | Score[c] | Acc. no.[d] | | % Similiar[e] |
|---|---|---|---|---|---|---|---|---|
| 2.56 | *6795-R1 | KC009247 | Mortierella hyalina | EDFL | 1205 | JN940867 | 100 | 99 |
| 2.56 | 6798-S2 | KC009253 | Penicillium roqueforti | Ascomycota | 946 | JQ434686 | 100 | 100 |
| 2.56 | 6799-S2 | KC009256 | Cordyceps militaris | Ascomycota | 428 | JQ286887 | 100 | 99 |
| 2.56 | *6799-R3 | KC009257 | Trichosporon dulcitum | Basidiomycota | 992 | AF444428 | 100 | 100 |
| 2.56 | *6801-S6 | KC009259 | Verticillium leptobactrum | Ascomycota | 907 | JQ410322 | 100 | 99 |
| 2.56 | 6802-S6 | KC009260 | Cuspidatispora xiphiago | Ascomycota | 874 | DQ376251 | 100 | 99 |
| 2.56 | 6803-S3 | KC009261 | Pochonia bulbillosa | Ascomycota | 918 | JQ780662 | 100 | 99 |
| 2.56 | 6804-R4 | KC009262 | Trichoderma viridescens | Ascomycota | 880 | HM535608 | 100 | 99 |
| 2.56 | 6805-S6 | KC009264 | Mammaria echinobotryoides | Ascomycota | 756 | DQ376251 | 100 | 99 |
| 2.56 | *6805-R1 | KC009265 | Mortierella alpina | EDFL | 680 | AB517932 | 100 | 99 |
| 2.56 | 6806-R3 | KC009269 | Oidiodendron tenuissimum | Ascomycota | 843 | AB040706 | 100 | 97 |
| 2.56 | *6807-S2 | KC009270 | Penicillium quercetorum | Ascomycota | 819 | AY443471 | 100 | 99 |
| 2.56 | *6807-S5 | KC009271 | Kernia pachypleura | Ascomycota | 472 | DQ318208 | 100 | 97 |
| 2.56 | 6809-S2 | KC009275 | Simplicillium sp. | Ascomycota | 662 | AB378540 | 99 | 95 |
| 5.13 | 6796-S3 | KC009248 | Arthroderma silverae | Ascomycota | 859 | AY176729 | 100 | 99 |
| 5.13 | *6796- | KC009249 | Mortierella parvispora | EDFL | 1182 | HM849689 | 100 | 99 |

| | S4 | | | | | | | |
|---|---|---|---|---|---|---|---|---|
| 5.13 | 6796-R1 | KC009250 | *Thamnidium elegans* | EDFL | 1162 | AB614353 | 100 | 99 |
| 5.13 | *6797-R1 | KC009252 | *Mortierella sarnyensis* | EDFL | 1229 | FJ161944 | 100 | 99 |
| 5.13 | *6817-S4 | KC009284 | *Mucor hiemalis* | EDFL | 1086 | JN315041 | 100 | 99 |
| 10.26 | 6799-R4 | KC009258 | *Coprinus callinus* | Ascomycota | 861 | AB470586 | 100 | 98 |
| 10.26 | 6805-S3 | KC009263 | *Chrysosporium vallenarense* | Ascomycota | 904 | AY176729 | 100 | 99 |
| 17.95 | 6808-S6 | KC009274 | *Kernia retardata* | Ascomycota | 933 | AB470603 | 98 | 99 |

[a]Relative abundance for the combined libraries, which was used to sort the entries
[b]OTUs were characterized by Mothur program [1], the OTU is ≥97% similar to a fungal isolate
[c]BLASTN [2] score value
[d]Accession number of the closest database match
[e]Level of similarity for pairwise alignments with the closest match, using the Martinez-Needleman-Wunsch algorithm [2]
*Common OTUs identified from ITS2, ITS, and LSU sequences

Table S7. Details of ITS sequences of fungal clones recovered by CI method

| Sum (%)[a] | OTU[b] Accession no. | Best BLAST hit | | | | | % Similiar[f] |
|---|---|---|---|---|---|---|---|
| | | Taxon | Phylum | Score[c] | Acc. no.[d] | Coverage[e] | |
| 0.41 | **S01-02 JX898552** | **Uncultured *Trichosporon* sp. clone** | **Basidiomycota** | **569** | **AB180201** | **74** | **99** |
| 0.41 | S01-05 JX898607 | Uncultured soil fungus clone | EDFL | 424 | JQ666464 | 47 | 92 |
| 0.41 | S01-15 JX898609 | Uncultured soil fungus clone | EDFL | 137 | EU517006 | 19 | 90 |
| 0.41 | S01-16 JX898610 | Uncultured soil fungus clone | EDFL | 292 | EF635761 | 32 | 96 |
| 0.41 | S06-36 JX898614 | Uncultured fungus clone | EDFL | 233 | EF392540 | 18 | 93 |
| 0.41 | S06-40 JX898596 | Uncultured *Hyaloraphidium* sp. clone | Chytridiomycota | 334 | AY997055 | 68 | 80 |
| 0.41 | S07-05 JX898597 | Uncultured *Sympodiomyces* sp. clone | Ascomycota | 520 | HQ623607 | 99 | 87 |
| 0.41 | S01-17 JX898611 | Uncultured *Mortierella* sp. clone | EDFL | 924 | HQ630349 | 100 | 96 |
| 0.41 | S01-19 JX898559 | Uncultured *Verrucariales* sp. clone | Ascomycota | 697 | HQ022024 | 75 | 91 |
| 0.41 | S02-02 JX605057 | Uncultured fungus clone | EDFL | 460 | FM178265 | 100 | 89 |
| 0.41 | S01-11 JX605055 | Uncultured *Trichosporon* sp. clone | Ascomycota | 767 | FJ943427 | 100 | 98 |
| 0.41 | S01-13 JX898608 | Uncultured soil fungus clone | EDFL | 169 | EU489943 | 15 | 90 |
| 0.41 | S02-07 JX898611 | Uncultured fungus clone | Ascomycota | 355 | JX974777 | 56 | 92 |

| 0.41 | S02-16 JX605066 | Uncultured *Debaryomyces sp.* clone | Ascomycota | 1046 | KC111444 | 100 | 100 |
| --- | --- | --- | --- | --- | --- | --- | --- |
| 0.41 | S02-17 JX898565 | Uncultured fungus clone | EDFL | 329 | FM178265 | 86 | 86 |
| 0.41 | S02-37 JX898572 | Uncultured *Trichosporon* sp. clone | Basidiomycota | 608 | HF558657 | 66 | 98 |
| 0.41 | S02-38 JX898573 | Uncultured *Mortierella* sp. clone | EDFL | 723 | HQ211948 | 99 | 89 |
| 0.41 | S04-18 JX898575 | Uncultured *Ganoderma* sp. clone clone | Basidiomycota | 654 | JN048773 | 75 | 99 |
| 0.41 | S04-24 JX605097 | Uncultured soil fungus clone | EDFL | 497 | JQ666392 | 100 | 93 |
| 0.41 | *S04-26 JX898576 | Uncultured *Penicillium sp.* clone | Ascomycota | 959 | JX270368 | 100 | 99 |
| 0.41 | S04-34 JX605106 | Uncultured soil fungus clone | EDFL | 545 | JQ666392 | 100 | 96 |
| 0.41 | S04-40 JX898577 | Uncultured Glomeromycota clone | Glomeromycota | 420 | GU392007 | 53 | 92 |
| 0.41 | *S05-02 JX898578 | Uncultured *Trichosporon dulcitum* clone | Basidiomycota | 815 | HM136684 | 100 | 99 |
| 0.41 | *S05-13 JX605117 | Uncultured *Thamnidium elegans* clone | EDFL | 1153 | AB113025 | 100 | 99 |
| 0.41 | S05-16 JX605120 | Uncultured *Chrysosporium* sp. clone | Ascomycota | 852 | AM949568 | 100 | 94 |
| 0.41 | S05-18 JX898582 | Uncultured Pseudeurotiaceae sp. clone | Ascomycota | 564 | JX270610 | 62 | 99 |
| 0.41 | S05-19 JX605122 | Uncultured *Geomyces* sp. clone | Ascomycota | 575 | HQ211533 | 100 | 89 |
| 0.41 | S05-31 JX898584 | Uncultured *Trichosporon* sp. clone | Basidiomycota | 619 | HF558657 | 66 | 99 |
| 0.41 | S06-09 JX898590 | Uncultured fungus clone | EDFL | 488 | EF635761 | 74 | 84 |
| 0.41 | S06-11 JX605145 | Uncultured Ascomycota clone | Ascomycota | 648 | HQ211979 | 92 | 96 |
| 0.41 | S06-12 | Uncultured fungus clone | Ascomycota | 808 | EF434057 | 97 | 88 |
| | JX898591 | | | | | | |
| 0.41 | S06-13 JX605146 | Uncultured fungus clone | EDFL | 484 | AJ920022 | 100 | 89 |
| 0.41 | *S06-15 JX605148 | Uncultured *Penicillium brevicompactum* | Ascomycota | 924 | HM469408 | 100 | 100 |
| 0.41 | S06-16 JX898613 | Uncultured soil fungus clone | Chytridiomycota | 257 | EU480336 | 26 | 94 |
| 0.41 | S11-04 JX898619 | Uncultured fungus clone | EDFL | 113 | JQ313110 | 18 | 84 |
| 0.41 | S11-06 JX898620 | Uncultured *Leptodiscella sp.* clone | EDFL | 127 | FR821312 | 17 | 83 |
| 0.41 | S11-08 JX898621 | Uncultured fungus clone | EDFL | 250 | HQ693513 | 25 | 90 |
| 0.41 | S11-09 JX898622 | Uncultured fungus clone | EDFL | 295 | EU825633 | 24 | 95 |
| 0.41 | S11-10 JX898604 | Uncultured fungus clone | EDFL | 358 | JQ681169 | 88 | 81 |
| 0.41 | S11-12 JX898623 | Uncultured fungus clone | EDFL | 134 | DQ421220 | 23 | 77 |
| 0.41 | S11-16 JX898624 | Uncultured fungus clone | EDFL | 291 | EU517006 | 28 | 97 |
| 0.41 | S11-20 JX675217 | Uncultured *Exophiala dermatitidis* clone | Ascomycota | 1059 | GU942305 | 100 | 99 |
| 0.41 | S09-15 JX898616 | Uncultured fungus clone | EDFL | 282 | HQ829352 | 26 | 90 |
| 0.41 | S09-19 JX898603 | Uncultured fungus clone | EDFL | 302 | HQ829349 | 68 | 87 |
| 0.41 | S11-01 JX605210 | Uncultured *Lycoperdon pyriforme* clone | Basidiomycota | 1157 | AY854075 | 100 | 99 |
| 0.41 | S11-02 JX898618 | Uncultured fungus clone | EDFL | 132 | EU718671 | 34 | 86 |
| 0.41 | S07-23 JX898600 | Uncultured *Mortierella* sp. Finse clone | EDFL | 733 | AJ541798 | 71 | 99 |
| 0.41 | S09-13 JX898615 | Uncultured soil fungus clone | EDFL | 197 | JQ666448 | 33 | 89 |
| 0.41 | S06-19 | Uncultured | Basidiomycota | 994 | EU030394 | 100 | 99 |
| | JX605151 | tremellomycete clone | | | | | |
| 0.41 | S06-22 JX898592 | Uncultured fungus clone | EDFL | 652 | EU516682 | 72 | 93 |
| 0.41 | S09-08 JX605203 | Uncultured fungus clone | Ascomycota | 668 | EU754961 | 100 | 94 |
| 0.83 | *S11-03 JX605211 | Uncultured *Ganoderma* sp. clone | Basidiomycota | 1000 | AF255097 | 100 | 99 |
| 0.83 | S01-01 JX605050 | Uncultured soil fungus clone | EDFL | 982 | GU083255 | 100 | 99 |
| 0.83 | S01-07 JX898554 | Uncultured *Mortierella* sp. clone | EDFL | 852 | JX270390 | 86 | 98 |
| 0.83 | S01-14 JX898557 | Uncultured Ascomycota clone | Ascomycota | 285 | HM240007 | 51 | 84 |
| 0.83 | S02-18 JX605067 | Uncultured *Mortierella hypsicladia* clone | EDFL | 1072 | HQ630302 | 100 | 98 |
| 0.83 | S05-25 JX605127 | Uncultured *Gymnoascus* sp. clone | Ascomycota | 1011 | JX270524 | 100 | 100 |
| 0.83 | S07-09 JX605168 | Uncultured *Paecilomyces inflatus* clone | Ascomycota | 868 | GU566291 | 100 | 99 |
| 0.83 | S06-23 JX605152 | Uncultured *Thamnidium elegans* clone | EDFL | 884 | AB113013 | 100 | 92 |
| 0.83 | S04-15 JX898612 | Uncultured fungus clone | EDFL | 298 | FJ528704 | 29 | 94 |
| 1.24 | S02-03 JX898562 | Uncultured *Gymnoascus* sp. clone | Ascomycota | 1002 | JX270593 | 100 | 99 |
| 1.24 | S09-03 JX898601 | Uncultured fungus clone | EDFL | 612 | EF635761 | 89 | 93 |
| 1.24 | *S06-17 JX605149 | Uncultured *Mortierella cf. gamsii* clone | EDFL | 1023 | HQ630307 | 100 | 100 |
| 1.24 | S06-33 JX898595 | Uncultured fungus clone | EDFL | 893 | EU516682 | 92 | 92 |
| 1.66 | *S01-06 JX605052 | Uncultured *Mortierella* sp. clone | EDFL | 1020 | JX270478 | 100 | 99 |
| 1.66 | S05-26 JX605128 | Uncultured fungus clone | Chytridiomycota | 488 | EU480016 | 100 | 84 |
| 2.07 | *S09-09 | Uncultured *Mortierella* | EDFL | 989 | GU327518 | 100 | 99 |

| | JX605204 | sp. clone | | | | | |
|---|---|---|---|---|---|---|---|
| 2.07 | *S02-21 JX605069 | Uncultured *Helicostylum pulchrum* clone | EDFL | 1099 | JQ319049 | 100 | 99 |
| 2.90 | *S05-30 JX605132 | Uncultured *Chaetomium* sp. clone | Ascomycota | 881 | GU934510 | 100 | 99 |
| 2.90 | *S06-14 JX605147 | Uncultured *Geomyces pannorum* clone | Ascomycota | 895 | JF311913 | 100 | 100 |
| 8.29 | *S02-19 JX898566 | Uncultured *Mortierella* sp. clone | EDFL | 1077 | JX270382 | 100 | 99 |
| 17.84 | *S04-33 JX605105 | Uncultured *Trichosporon dulcitum* clone | Basidiomycota | 859 | HF558657 | 100 | 100 |
| 26.97 | *S07-15 JX605174 | Uncultured *Mortierella* sp. clone | EDFL | 1058 | AJ541798 | 100 | 99 |

[a]Relative abundance for the combined libraries, which was used to sort the entries
[b]OTUs were characterized by Mothur platform [1], the OTU is ≥97% similar to a fungal isolate
[c]BLASTN [2] score value
[d]Accession number of the closest database match
[e]Coverage of pairwise alignment of the closest database match
[f]Level of similarity for pairwise alignments with the closest match, using the Martinez-Needleman-Wunsch algorithm [2]
*Common OTUs recovered from LSU and ITS clone libraries

Table S8. Details of LSU sequences of fungal clones recovered by CI method

| Sum(%)[a] | OTU[b] | Accession no. | Best BLAST hit Taxon | Phylum | Score[c] | Acc. no.[d] | % Similiar[e] |
|---|---|---|---|---|---|---|---|
| 0.53 | D06-36 | JX534740 | Uncultured *Trichosporon sp.* clone | Basidiomycota | 767 | KF212338 | 99 |
| 0.53 | D09-09 | JX534814 | Uncultured *Trichosporon sp.* clone | Basidiomycota | 669 | JN939493 | 69 |
| 0.53 | D12-35 | JX534928 | Uncultured soil fungus clone | EDFL | 1024 | KF565789 | 95 |
| 0.53 | D09-31 | JX534833 | Uncultured soil fungus clone | Ascomycota | 1081 | KF566345 | 94 |
| 0.53 | D08-30 | JX534799 | Uncultured soil fungus clone | EDFL | 1003 | EU516994 | 93 |
| 0.53 | D09-27 | JX534829 | Uncultured soil fungus clone | Ascomycota | 939 | KF565913 | 87 |
| 0.53 | *D03-08 | JX534651 | Uncultured *Mortierella hyalina* clone | EDFL | 1173 | JN940868 | 98 |
| 0.53 | D08-13 | JX534785 | Uncultured soil fungus clone | EDFL | 603 | KF566652 | 99 |
| 0.53 | D03-23 | JX534664 | Uncultured *M. exigua voucher* clone | EDFL | 1243 | FJ161943 | 99 |
| 0.53 | D03-40 | JX534677 | Uncultured *Trichosporon* clone | Basidiomycota | 798 | JN939493 | 93 |
| 0.53 | D09-15 | JX534820 | Uncultured soil fungus clone | EDFL | 575 | KF565528 | 66 |
| 0.53 | D04-10 | JX534682 | Uncultured *Trichosporon sp.* clone | Basidiomycota | 630 | JN939493 | 93 |
| 0.53 | D07-06 | JX534747 | Uncultured *Guehomyces pullulans* clone | Basidiomycota | 981 | GQ202976 | 100 |
| 0.53 | D10-02 | JX534840 | Uncultured *Hypochniciellum molle* clone | Basidiomycota | 992 | GU187667 | 99 |
| 0.53 | D01-07 | JX534607 | Uncultured *Mortierella sp.* clone | EDFL | 1009 | JX976166 | 94 |
| 0.53 | D01-34 | JX534625 | Uncultured *Mortierella sp.* clone | EDFL | 780 | HQ667424 | 72 |
| 0.53 | D03-03 | JX534647 | Uncultured *Trichosporon sp.* clone | Basidiomycota | 579 | JN939493 | 66 |
| 0.53 | D01-04 | JX534605 | Uncultured *Doratomyces stemonitis* clone | Ascomycota | 933 | DQ836907 | 99 |
| 0.53 | D09-06 | JX534812 | Uncultured soil fungus clone | EDFL | 881 | KF566043 | 90 |
| 0.53 | D06-25 | JX534735 | Uncultured *Basidiomycota* clone | Basidiomycota | 841 | DQ341804 | 95 |
| 0.53 | D03-26 | JX534666 | Uncultured *Mortierella sp.* clone | EDFL | 1122 | JN940866 | 96 |
| 0.53 | D03-01 | JX534646 | Uncultured *Mortierella sp.* clone | EDFL | 915 | KC018415 | 92 |
| 0.53 | D03-06 | JX534649 | Uncultured *Mortierella sp.* clone | EDFL | 1081 | KC018408 | 96 |
| 0.53 | D05-24 | JX534711 | Uncultured *Mortierella sp.* clone | EDFL | 822 | FJ161938 | 88 |
| 0.53 | D11-25 | JX534895 | Uncultured *Coprinellus micaceus* clone | Basidiomycota | 1005 | AY207182 | 99 |
| 0.53 | *D05-04 | JX534701 | Uncultured *Mortierella indohii* clone | EDFL | 1149 | EU736318 | 98 |
| 0.53 | D11-21 | JX534892 | Uncultured *Mortierella polycephala* clone | EDFL | 1192 | KC018297 | 97 |
| 0.53 | D09-38 | JX534837 | Uncultured *Trichosporon sp.* clone | Basidiomycota | 599 | JN939493 | 65 |
| 0.53 | D05-16 | JX534707 | Uncultured *Trichosporon middelhovenii* clone | Basidiomycota | 937 | AB180198 | 99 |
| 0.53 | D09-29 | JX534831 | Uncultured *Fusarium sp.* clone | Ascomycota | 486 | EU860057 | 62 |
| 0.53 | *D03-33 | JX534672 | Uncultured *Mortierella gamsii* clone | EDFL | 1236 | HQ667384 | 100 |
| 0.53 | D10-16 | JX534853 | Uncultured *Mortierella polycephala* clone | EDFL | 1212 | JN939145 | 99 |
| 0.53 | D01-21 | JX534617 | Uncultured soil fungus clone | EDFL | 612 | KC965494 | 88 |
| 0.53 | D05-40 | JX534724 | Uncultured soil fungus clone | Ascomycota | 529 | KJ150290 | 82 |
| 0.53 | D01-03 | JX534604 | Uncultured *Gyromitus sp.* clone | Ascomycota | 411 | EF681955 | 68 |
| 0.53 | D12-38 | JX534930 | Uncultured *Pluteus sp.* clone | Basidiomycota | 634 | HM562242 | 89 |
| 0.53 | D12-36 | JX534929 | Uncultured soil fungus clone | Ascomycota | 806 | JX043247 | 88 |
| 0.53 | D08-35 | JX534804 | Uncultured soil fungus clone | EDFL | 599 | EU489955 | 94 |
| 0.53 | D06-01 | JX534725 | Uncultured soil fungus clone | EDFL | 656 | EU489955 | 93 |
| 0.53 | D11-15 | JX534888 | Uncultured soil fungus clone | EDFL | 475 | GU376391 | 70 |
| 0.53 | D05-30 | JX534717 | Uncultured *Gymnascella sp.* clone | Ascomycota | 566 | KC009164 | 64 |
| 0.53 | D08-16 | JX534788 | Uncultured soil fungus clone | EDFL | 662 | DQ901000 | 69 |
| 0.53 | D11-33 | JX534899 | Uncultured soil fungus clone | EDFL | 590 | JQ311729 | 88 |
| 0.53 | D02-07 | JX534632 | Uncultured *Geomyces destructans* clone | Ascomycota | 922 | KF212357 | 99 |
| 0.53 | D09-22 | JX534822 | Uncultured *Maunachytrium sp.* clone | Chytridiomycota | 630 | EF432822 | 86 |
| 0.53 | D09-05 | JX534811 | Uncultured *Chytridiales sp.* clone | Chytridiomycota | 472 | EF443139 | 81 |
| 0.53 | D12-26 | JX534922 | Uncultured *Saccharomyces cerevisiae* clone | Ascomycota | 961 | KJ506733 | 99 |
| 0.53 | D11-04 | JX534880 | Uncultured *Arthrographis sp.* clone | Ascomycota | 741 | AB116543 | 93 |
| 0.53 | D10-19 | JX534856 | Uncultured *Acremonium sp.* clone | Ascomycota | 549 | KJ194116 | 96 |
| 0.53 | D11-34 | JX534900 | Uncultured soil fungus clone | EDFL | 470 | JX898660 | 89 |
| 0.53 | D01-14 | JX534611 | Uncultured soil fungus clone | EDFL | 627 | DQ900984 | 80 |
| 0.53 | D02-01 | JX534627 | Uncultured soil fungus clone | EDFL | 928 | JQ310857 | 95 |
| 0.53 | *D02-03 | JX534628 | Uncultured *Mortierella indohii* clone | EDFL | 1315 | EU688966 | 99 |
| 0.53 | D06-27 | JX534737 | Uncultured soil fungus clone | Ascomycota | 1162 | KF566486 | 97 |
| 0.53 | D06-30 | JX534739 | Uncultured soil fungus clone | Ascomycota | 652 | JQ311054 | 75 |
| 0.53 | *D10-01 | JX534839 | Uncultured *Mortierella indohii* clone | EDFL | 1186 | KC018415 | 99 |
| 0.53 | D09-28 | JX534830 | Uncultured *Zygoascus sp.* clone | Ascomycota | 693 | HM036077 | 88 |
| 0.53 | D02-37 | JX534643 | Uncultured soil fungus clone | EDFL | 977 | KC965607 | 96 |
| 0.53 | D08-24 | JX534795 | Uncultured soil fungus clone | EDFL | 939 | JQ311523 | 85 |
| 0.53 | D02-38 | JX534644 | Uncultured soil fungus clone | EDFL | 512 | JX534845 | 65 |

| 0.53 | D09-23 | JX534825 | Uncultured *Pichia sp.* clone | Ascomycota | 787 | HM036077 | 92 |
|---|---|---|---|---|---|---|---|
| 0.53 | D05-31 | JX534718 | Uncultured *Torrubiella wallacei* clone | Ascomycota | 856 | AY184967 | 97 |
| 0.53 | D02-05 | JX534630 | Uncultured soil fungus clone | EDFL | 442 | KF565861 | 64 |
| 0.53 | D01-29 | JX534621 | Uncultured soil fungus clone | EDFL | 590 | JX545259 | 88 |
| 0.53 | *D10-24 | JX534860 | Uncultured *Penicillium sp.* clone | Ascomycota | 939 | KF880926 | 99 |
| 0.53 | D10-25 | JX534861 | Uncultured *Neogymnomyces sp.* clone | Ascomycota | 867 | AY176716 | 97 |
| 0.53 | D10-30 | JX534866 | Uncultured *Acremonium charticola* clone | Ascomycota | 918 | HQ232015 | 99 |
| 0.53 | D10-38 | JX534874 | Uncultured *Chrysosporium sulfureum* clone | Ascomycota | 950 | KC989731 | 99 |
| 0.53 | D09-02 | JX534809 | Uncultured soil fungus clone | EDFL | 992 | EU861720 | 90 |
| 0.53 | D04-27 | JX534690 | Uncultured soil fungus clone | EDFL | 658 | EU861720 | 82 |
| 0.53 | D06-29 | JX534738 | Uncultured soil fungus clone | EDFL | 833 | JQ311621 | 82 |
| 0.53 | D10-39 | JX534875 | Uncultured *Doratomyces sp.* clone | Ascomycota | 675 | DQ836907 | 90 |
| 0.53 | D08-37 | JX534805 | Uncultured *Fusarium sp.* clone | Ascomycota | 830 | EU860057 | 95 |
| 0.53 | *D07-37 | JX534773 | Uncultured *Geomyces pannorum* clone | Ascomycota | 909 | GU951697 | 99 |
| 0.53 | D08-05 | JX534779 | Uncultured *Geomyces destructans* clone | Ascomycota | 911 | KC171321 | 98 |
| 0.53 | D05-22 | JX534709 | Uncultured *Chrysosporium merdarium* clone | Ascomycota | 665 | KC989728 | 90 |
| 0.53 | D05-27 | JX534714 | Uncultured *Amauroascus sp.* clone | Ascomycota | 865 | AB075324 | 97 |
| 0.53 | D05-01 | JX534698 | Uncultured *Geomyces pannorum* clone | Ascomycota | 835 | GU951694 | 97 |
| 0.53 | D03-32 | JX534671 | Uncultured *Cercophora sparsa* clone | Ascomycota | 856 | AY587937 | 97 |
| 0.53 | D07-22 | JX534761 | Uncultured *Oidiodendron sp.* | Ascomycota | 893 | AB040706 | 98 |

| | | | clone | | | | |
|---|---|---|---|---|---|---|---|
| 0.53 | D01-01 | JX534602 | Uncultured *Mortierella sp.* clone | EDFL | 773 | KC018389 | 73 |
| 0.53 | D02-16 | JX534635 | Uncultured *Leuconeurospora sp.* clone | Ascomycota | 839 | AF096193 | 96 |
| 0.53 | D10-20 | JX534857 | Uncultured *Oidiodendron sp.* clone | Ascomycota | 736 | AB040706 | 93 |
| 0.53 | D10-31 | JX534867 | Uncultured *Acremonium sp.* clone | Ascomycota | 806 | HQ231987 | 96 |
| 0.53 | D11-17 | JX534889 | Uncultured *Calycellina populina* clone | Ascomycota | 928 | JN086693 | 98 |
| 0.53 | D07-11 | JX534751 | Uncultured *Cephalotheca sp.* clone | Ascomycota | 616 | FJ808681 | 66 |
| 0.53 | D07-20 | JX534759 | Uncultured *Cephalotheca sulfurea* clone | Ascomycota | 852 | FJ808681 | 97 |
| 0.53 | D07-17 | JX534756 | Uncultured *Cephalotheca sulfurea* clone | Ascomycota | 883 | KC311468 | 98 |
| 0.53 | D07-35 | JX534772 | Uncultured *Cephalotheca sp.* clone | Ascomycota | 885 | FJ808681 | 98 |
| 0.53 | D08-38 | JX534806 | Uncultured *Doratomyces sp.* clone | Ascomycota | 876 | KC009280 | 97 |
| 0.53 | D09-11 | JX534816 | Uncultured *Chalara sp.* clone | Ascomycota | 625 | FJ176257 | 88 |
| 0.53 | D10-37 | JX534873 | Uncultured *Doratomyces sp.* clone | Ascomycota | 863 | AB470586 | 98 |
| 0.53 | D08-34 | JX534803 | Uncultured *Chalara sp.* clone | Ascomycota | 774 | FJ176257 | 93 |
| 0.53 | D06-09 | JX534729 | Uncultured *Anisomeridium sp.* clone | Ascomycota | 791 | DQ782906 | 95 |
| 0.53 | D03-38 | JX534675 | Uncultured *Arthrobotrys superba* clone | Ascomycota | 931 | EF445988 | 99 |
| 0.53 | D02-39 | JX534645 | Uncultured *Chalara sp.* clone | Ascomycota | 802 | FJ176257 | 94 |
| 0.53 | D02-20 | JX534637 | Uncultured *Chalara sp.* clone | Ascomycota | 689 | FJ176257 | 91 |
| 0.53 | D09-07 | JX534813 | Uncultured *Kotlabaea sp.* clone | Ascomycota | 926 | DQ220356 | 99 |

| 0.53 | D11-23 | JX534894 | Uncultured *Cosmospora viliuscula* clone | Ascomycota | 935 | KC291785 | 99 |
| --- | --- | --- | --- | --- | --- | --- | --- |
| 0.53 | D12-03 | JX534907 | Uncultured *Acremonium aff. curvulum* clone | Ascomycota | 931 | HQ232031 | 99 |
| 0.53 | D04-05 | JX534680 | Uncultured *Kotlabaea sp.* clone | Ascomycota | 922 | DQ220356 | 98 |
| 0.53 | D03-19 | JX534661 | Uncultured *Heydenia alpina* clone | Ascomycota | 907 | HQ596526 | 100 |
| 0.53 | D07-32 | JX898649 | Uncultured *Cephalotheca sp.* clone | Ascomycota | 555 | KC311468 | 65 |
| 0.53 | D09-03 | JX898653 | Uncultured soil fungus clone | Glomeromycota | 375 | KF566643 | 55 |
| 0.53 | D11-14 | JX898660 | Uncultured soil fungus clone | Ascomycota | 414 | JX067943 | 70 |
| 0.53 | D11-30 | JX898664 | Uncultured soil fungus clone | EDFL | 279 | KF651080 | 61 |
| 0.53 | D09-18 | JX898654 | Uncultured soil fungus clone | EDFL | 436 | KF568205 | 89 |
| 0.53 | D06-14 | JX898644 | Uncultured soil fungus clone | Glomeromycota | 217 | JN049542 | 91 |
| 0.53 | D06-24 | JX898645 | Uncultured *Trichosporon sp.* clone | Basidiomycota | 616 | JN939493 | 66 |
| 0.53 | D09-37 | JX898656 | *Uncultured soil fungus clone* | Chytridiomycota | 285 | EU379177 | 91 |
| 0.53 | D08-26 | JX898651 | *Uncultured soil fungus clone* | EDFL | 392 | KF568432 | 87 |
| 0.53 | D08-27 | JX898652 | *Uncultured Microascus sp. clone* | Ascomycota | 510 | KC009283 | 94 |
| 0.53 | D08-03 | JX898650 | Uncultured soil fungus clone | Ascomycota | 257 | GU928603 | 80 |
| 0.53 | D09-19 | JX898655 | Uncultured soil fungus clone | Ascomycota | 180 | KF738159 | 87 |
| 0.53 | D10-05 | JX898657 | Uncultured soil fungus clone | Ascomycota | 137 | GQ144684 | 82 |
| 0.53 | D11-10 | JX898659 | Uncultured soil fungus clone | Basidiomycota | 195 | AY752989 | 88 |
| 0.53 | D11-16 | JX898662 | Uncultured soil fungus clone | Basidiomycota | 182 | KC176336 | 87 |
| 0.53 | D11-26 | JX898663 | Uncultured soil fungus clone | EDFL | 99 | KF750512 | 88 |
| 0.53 | D11-29 | JX898664 | Uncultured soil fungus clone | EDFL | 388 | KF568432 | 58 |
| 0.53 | D11-37 | JX898666 | Uncultured soil fungus clone | EDFL | 368 | KF567069 | 55 |
| 0.53 | D12-27 | JX898674 | Uncultured soil fungus clone | Glomeromycota | 211 | JN937441 | 90 |
| 0.53 | D12-32 | JX898675 | Uncultured soil fungus clone | Basidiomycota | 187 | KF567090 | 87 |
| 0.53 | D02-40 | JX898641 | Uncultured *Calcarisporiella sp.* | EDFL | 407 | AB617740 | 88 |

| | | | clone | | | | |
|---|---|---|---|---|---|---|---|
| 0.53 | D01-39 | JX898638 | Uncultured soil fungus clone | EDFL | 99 | KF750512 | 71 |
| 0.53 | D06-33 | JX898647 | Uncultured soil fungus clone | EDFL | 403 | AB617740 | 87 |
| 0.53 | D07-26 | JX898649 | Uncultured soil fungus clone | EDFL | 399 | KF568432 | 50 |
| 0.53 | D06-13 | JX898644 | Uncultured *Pseudogymnoascus sp.* clone | Ascomycota | 538 | KF017871 | 64 |
| 0.53 | D01-19 | JX534615 | Uncultured soil fungus clone | EDFL | 442 | JX067943 | 88 |
| 0.53 | *D06-04 | JX534727 | Uncultured *Mortierella exigua* clone | EDFL | 1218 | KC009154 | 99 |
| 0.53 | D09-13 | JX534818 | Uncultured *Mortierella zonata* clone | EDFL | 1242 | KC018434 | 99 |
| 0.53 | D05-38 | JX898642 | Uncultured soil fungus clone | EDFL | 243 | EF681910 | 85 |
| 0.53 | D06-02 | JX898643 | Uncultured soil fungus clone | EDFL | 207 | DQ393450 | 86 |
| 0.53 | D10-06 | JX534843 | Uncultured *Hypochniciellum sp.* clone | Basidiomycota | 795 | AY586679 | 93 |
| 0.53 | D10-29 | JX534865 | Uncultured *Amylocorticiales sp.* clone | Basidiomycota | 1007 | KC514898 | 99 |
| 0.53 | D12-08 | JX534910 | Uncultured soil fungus clone | EDFL | 1033 | FJ176706 | 95 |
| 0.53 | D12-28 | JX534923 | Uncultured soil fungus clone | EDFL | 771 | JX043247 | 87 |
| 0.53 | D12-13 | JX534911 | Uncultured soil fungus clone | EDFL | 795 | JX043247 | 88 |
| 0.53 | D12-14 | JX534912 | Uncultured *Hyphodontia sp.* clone | Basidiomycota | 985 | DQ340352 | 99 |
| 0.53 | D12-15 | JX534913 | Uncultured soil fungus clone | EDFL | 804 | JX043247 | 88 |
| 0.53 | D09-36 | JX534836 | Uncultured zygomycete clone | EDFL | 562 | EU490131 | 82 |
| 0.53 | D12-20 | JX534916 | Uncultured *Phialophora sp.* clone | Ascomycota | 972 | AB190421 | 99 |
| 0.53 | D12-22 | JX534918 | Uncultured *Phialophora dancoi* clone | Ascomycota | 942 | AB190421 | 100 |
| 0.53 | D09-25 | JX534827 | Uncultured zygomycete clone | EDFL | 569 | EU490131 | 83 |
| 0.53 | D11-03 | JX534879 | Uncultured Ascomycota clone | Ascomycota | 953 | HQ432987 | 98 |
| 0.53 | *D11-02 | JX534878 | Uncultured *Ganoderma australe* clone | Basidiomycota | 1005 | JN048792 | 100 |
| 0.53 | D11-06 | JX534882 | Uncultured Ascomycota clone | Ascomycota | 961 | HQ432987 | 99 |
| 0.53 | D11-12 | JX534886 | Uncultured soil fungus clone | Ascomycota | 931 | HQ432987 | 96 |
| 0.53 | D11-27 | JX534896 | Uncultured *Ganoderma sp.* clone | Basidiomycota | 992 | JN048792 | 99 |
| 0.53 | D07-23 | JX534762 | Uncultured *Fusarium sp.* clone | Ascomycota | 918 | EU860057 | 99 |
| 0.53 | D07-14 | JX534754 | Uncultured soil fungus clone | Ascomycota | 953 | KF566557 | 99 |
| 0.53 | D07-24 | JX534763 | Uncultured soil fungus clone | Ascomycota | 907 | KC557359 | 97 |
| 0.53 | D10-22 | JX534858 | Uncultured *Mortierella polycephala* clone | EDFL | 1223 | JN939146 | 99 |
| 0.53 | D12-33 | JX534926 | Uncultured *Phialophora sp.* clone | Ascomycota | 937 | AB190421 | 98 |
| 0.53 | *D11-05 | JX534881 | Uncultured *Ganoderma sp.* clone | Basidiomycota | 987 | JN048792 | 99 |
| 0.53 | D12-07 | JX534909 | Uncultured *Phialophora dancoi* clone | Ascomycota | 961 | AB190421 | 99 |
| 0.53 | D03-07 | JX534650 | Uncultured *Cercophora sp.* clone | Ascomycota | 909 | AY587937 | 99 |
| 0.53 | D03-22 | JX534663 | Uncultured *Cercophora sparsa* clone | Ascomycota | 924 | AY587937 | 98 |
| 0.53 | D03-29 | JX534668 | Uncultured *Cercophora sparsa* clone | Ascomycota | 929 | AY587937 | 100 |
| 0.53 | D02-32 | JX534641 | Uncultured *Leuconeurospora sp.* clone | Ascomycota | 928 | FJ176884 | 99 |
| 0.53 | D12-09 | JX898668 | Uncultured soil fungus clone | EDFL | 551 | KF566381 | 90 |
| 0.53 | D07-40 | JX534775 | Uncultured soil fungus clone | Ascomycota | 826 | KF567483 | 94 |
| 1.06 | D07-01 | JX534742 | Uncultured soil fungus clone | Ascomycota | 985 | KF566557 | 99 |
| 1.06 | D12-12 | JX898771 | Uncultured soil fungus clone | Glomeromycota | 545 | KF566381 | 99 |
| 1.06 | D12-19 | JX898773 | Uncultured soil fungus clone | EDFL | 556 | KF566381 | 91 |
| 1.06 | D03-15 | JX534658 | Uncultured *Mortierella indohii* clone | EDFL | 1194 | EU688966 | 99 |
| 1.06 | D03-18 | JX534660 | Uncultured *Tetracladium sp.* clone | Ascomycota | 935 | KF768462 | 99 |
| 1.06 | D12-06 | JX534908 | Uncultured *Cladosporium coralloides* clone | Ascomycota | 965 | KF611802 | 99 |
| 1.06 | D10-15 | JX534852 | Uncultured *Uncinocarpus reesii* clone | Ascomycota | 880 | JQ434632 | 98 |
| 1.06 | D03-10 | JX534653 | Uncultured *Fusarium sp.* clone | Ascomycota | 887 | GU055596 | 100 |
| 1.06 | D10-27 | JX534863 | Uncultured *Mortierella polycephala* clone | EDFL | 1223 | JN939145 | 99 |
| 1.06 | D01-20 | JX534616 | Uncultured *Leuconeurospora sp.* clone | Ascomycota | 957 | FJ176884 | 100 |
| 1.59 | D12-18 | JX534915 | Uncultured soil fungus clone | EDFL | 977 | EU516968 | 94 |
| 2.12 | D09-12 | JX534817 | Uncultured soil fungus clone | EDFL | 1083 | EU861720 | 92 |
| 2.65 | D01-06 | JX898636 | Uncultured soil fungus clone | Basidiomycota | 414 | JX898661 | 80 |
| 2.65 | D06-06 | JX534728 | Uncultured soil fungus clone | EDFL | 1151 | KF566486 | 97 |
| 3.18 | D01-31 | JX534622 | Uncultured soil fungus clone | EDFL | 1201 | KF565913 | 97 |
| 3.71 | D02-06 | JX534631 | Uncultured *Sphaerosporium equinum* clone | Ascomycota | 950 | JQ434638 | 100 |
| 3.71 | D11-07 | JX534883 | Uncultured *Oxyporus corticola* clone | Basidiomycota | 1022 | KC176679 | 100 |
| 3.71 | D02-28 | JX534640 | Uncultured *Mucor flavus* clone | EDFL | 1127 | EU071390 | 99 |
| 4.24 | D07-03 | JX534744 | Uncultured *Cephalotheca sulfurea* clone | Ascomycota | 948 | FJ808681 | 99 |
| 4.24 | D08-06 | JX534780 | Uncultured soil fungus clone | EDFL | 848 | KF565875 | 95 |
| 5.3 | D02-36 | JX534642 | Uncultured *Mortierella elongata* clone | EDFL | 1216 | FJ161938 | 99 |
| 5.83 | D06-03 | JX534726 | Uncultured *Pichia sp.* clone | Ascomycota | 918 | HM036077 | 97 |
| 7.42 | *D05-02 | JX534699 | Uncultured *Chaetomidium arxii* clone | Ascomycota | 852 | JX280740 | 98 |
| 7.95 | *D03-11 | JX534654 | Uncultured *Mortierella indohii* clone | EDFL | 1190 | EU736318 | 99 |

| 8.48 | D01-05 | JX534606 | Uncultured soil fungus clone | EDFL | 468 | GU376405 | 90 |
| 8.48 | D01-15 | JX534612 | Uncultured *Doratomyces sp.* clone | Ascomycota | 955 | KC009280 | 100 |
| 9.01 | *D01-18 | JX534614 | Uncultured *Geomyces pannorum* clone | Ascomycota | 957 | JQ768405 | 100 |
| 14.31 | *D03-05 | JX534648 | Uncultured *Trichosporon dulcitum* clone | Basidiomycota | 959 | JN939493 | 100 |

[a]Relative abundance for the combined libraries, which was used to sort the entries
[b]OTUs were characterized by Mothur program [1], the OTU is ≥97% similar to a fungal isolate
[c]BLASTN [2] score value
[d]Accession number of the closest database match
[e]Level of similarity for pairwise alignments with the closest match, using the Martinez-Needleman-Wunsch algorithm [2]
*Common OTUs recovered from LSU and ITS cloned libraries

Table S9. List of common OTUs identified by CD and CI methods

| Culture-dependent | | | Culture-independent | | | Best BLAST hit | | | | |
|---|---|---|---|---|---|---|---|---|---|---|
| Strain code [a] | ITS/LSU | Abundance [b] | Clone code [a] | ITS/LSU | Abundance [b] | Taxon | Phylum | Score [c] | Accession no. [d] | Similarity (%) |
| 6724-S7 KC009184 | LSU | 16 | D08-05 JX534778 | LSU | 11 | *Pseudogymnoascus destructans* | Ascomycota | 444/524 | GU944945 | 99-100 |
| 6732-R4 KC009199 | LSU | 13 | D04-13 JX534682 | LSU | 27 | *Trichosporon dulcitum* | Basidiomycota | 996/993 | JN939493 | 100 |
| 38252-S2 KC009141 | LSU | 14 | D05-01 JX534697 | LSU | 1 | *Geomyces pannorum* | Ascomycota | 961/835 | GU951689/GU951691 | 98-100 |
| 38275-R2 KC009340 | ITS | 11 | S02-21 JX675069 | ITS | 43 | *Helicostylum pulchrum* | EDL | 704/1076 | AB614353 | 99-100 |
| 39141-S3 KC009341 | ITS | 53 | S12-06 JX534908 | LSU | 2 | *Cladosporium cladosporioides* | Ascomycota | 524/525 | JX077073/JQ388759 | 99-100 |
| 39144-S5 KC009348 | ITS2 | 53 | S06-15 JX675148 | ITS | 1 | *Penicillium brevicompactum* | Ascomycota | 625/814 | AY37398/HM469408 | 99-100 |
| 6730-R5 KC009412 | ITS | 13 | D07-23 JX534761 | LSU | 3 | *Fusarium merismoides* | Ascomycota | 535/918 | EU860057 | 99-100 |
| 6747-R4 KC009441 | ITS | 30 | D07-06 JX534746 | LSU | 1 | *Guehomyces pullulans* | Basidiomycota | 682/531 | AF444417/FR774553 | 99-100 |
| 41557-1 KC009376 | ITS2 | 5 | D01-04 JX534605 | LSU | 1 | *Doratomyces stemonitis* | Ascomycota | 518/516 | JN104543/DQ836907 | 99-100 |
| 38273-R2 KC009334 | ITS | 2 | S05-30 JX675132 | ITS | 7 | *Chaetomium* sp. | Ascomycota | 854/486 | GU934510 | 99-100 |
| 6793-R1 KC009243 | LSU | 14 | S05-13 JX675117 | ITS | 1 | *Thamnidium elegans* | EDFL | 1092/630 | AB614353/AB113025 | 99 |
| 6734-R6 KC008918 | ITS | 3 | S06-17 JX675149 | ITS | 3 | *Mortierella cf. gamsii* | EDFL | 989/1023 | HQ630307 | 100 |
| 6794-S4 KC009029 | ITS | 13 | S02-16 JX675066 | ITS | 1 | *Debaryomyces hansenii* | Ascomycota | 566/1021 | HE681099 | 100 |

[a] One representative strain (from the CD method) and one representative clone (from the CI method) is presented for the 13 shared OTUs. [b] Abundance indicates the number of strains or clones for each OTU in the entire data set. [c] BLASTN score. [d] Accession number of the closest database match. EDFL, early diverging fungal lineages.

Table S10. Alpha diversity analyses of fungal DNA clones recovered in the WNS-infested environmental samples.

| Sample ID | Reads | Cut off = 0.03 | | | | | |
|---|---|---|---|---|---|---|---|
| | | OTU | ACE | Chao-1 | Coverage | Shannon | Simpson |
| LSU | 353 | 189 | 938 (783, 1134) | 595 (430, 873) | 0.611898 | 4.86 (4.75, 4.98) | 0.0101 (0.0071, 0.013) |
| ITS | 241 | 73 | 394 (298, 530) | 237 (140, 466) | 0.796680 | 3.04 (2.83, 3.25) | 0.115 (0.0874, 0.1427) |

Note: LSU, large-subunit rDNA; ITS, internal transcribed spacer rDNA.

Fig. S1. Rank-abundance plots for the culture-dependent and -independent investigations. Abundance of each OTU was indicated by the number of isolates in the culture-dependent approach (51 OTUs were singletons in 73 ITS OTUs) or the number of clones in the culture-independent approach (167 OTUs were singletons in 189 LUS OTUs).

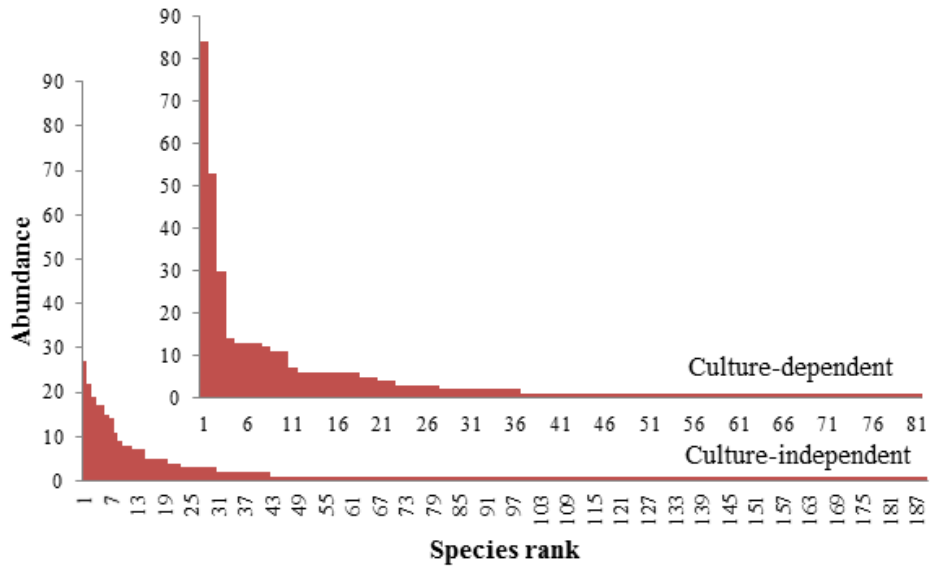

**Reference**

1. Schloss PD, Westcott SL, Ryabin T, Hall JR, Hartmann M, et al. (2009) Introducing mothur: open-source, platform-independent, community-supported software for describing and comparing microbial communities. Appl Environ Microbiol 75: 7537-7541.
2. Altschul SF, Madden TL, Schaffer AA, Zhang J, Zhang Z, et al. (1997) Gapped BLAST and PSI-BLAST: a new generation of protein database search programs. Nucleic Acids Res 25: 3389-3402.